\documentclass[sigconf,screen]{acmart}
\AtBeginDocument{%
  }

\setcopyright{acmlicensed}
\copyrightyear{2026}
\acmYear{2026}
\setcopyright{cc}
\setcctype{by}
\acmConference[CHI '26]{Proceedings of the 2026 CHI Conference on Human Factors in Computing Systems}{April 13--17, 2026}{Barcelona, Spain}
\acmBooktitle{Proceedings of the 2026 CHI Conference on Human Factors in Computing Systems (CHI '26), April 13--17, 2026, Barcelona, Spain}
\acmPrice{}
\acmDOI{10.1145/3772318.3791803}
\acmISBN{979-8-4007-2278-3/2026/04}

\usepackage[ruled,vlined]{algorithm2e}
\usepackage{enumitem}
\usepackage{hyperref}

\usepackage{marvosym} 

\makeatletter
\renewcommand*\@fnsymbol[1]{%
  \ifcase#1\or
    \ensuremath{\dagger}
  \or
    \Letter
  \or
    \ensuremath{\ddagger}
  \or
    \S
  \or
    \P
  \or
    \|
  \or
    **
  \or
    \ensuremath{\dagger\dagger}
  \or
    \Letter\Letter
  \or
    \ensuremath{\ddagger\ddagger}
  \else
    \@ctrerr
  \fi}
\makeatother

\usepackage{listings}
\usepackage{xcolor}
\usepackage{float}   

\usepackage{soul}
\newif\ifshowchanges
\showchangestrue   

\ifshowchanges
    \newcommand{\delBWT}[1]{\textcolor{blue}{\st{#1}}} 
    \newcommand{\delMYX}[1]{\textcolor{red}{\st{#1}}}  
    \newcommand{\delCQX}[1]{\textcolor{green}{\st{#1}}} 
\else
    \newcommand{\delBWT}[1]{}       
    
    \newcommand{\delMYX}[1]{}
    
    \newcommand{\delCQX}[1]{}
\fi

\begin{document}

\title{Roomify: Spatially-Grounded Style Transformation for Immersive Virtual Environments}



\author{Xueyang Wang}
\authornote{Co-first authors.}
\orcid{0000-0002-9797-9491}
\email{wang-xy22@mails.tsinghua.edu.cn}
\affiliation{
    \institution{Tsinghua University}
    \city{Beijing}
    \country{China}
}

\author{Qinxuan Cen}
\authornotemark[1] 
\orcid{0009-0008-0076-3151}
\email{cenqinxuan@bupt.edu.cn}
\affiliation{
    \institution{Beijing University of Posts and Telecommunications}
    \city{Beijing}
    \country{China}
}

\author{Weitao Bi}
\orcid{0009-0009-5856-5067}
\email{bwt24@mails.tsinghua.edu.cn}
\affiliation{
    \institution{Tsinghua University}
    \city{Beijing}
    \country{China}
}

\author{Yunxiang Ma}
\orcid{0009-0001-2289-9673}
\email{yunxianm@andrew.cmu.edu}
\affiliation{
    \institution{Carnegie Mellon University}
    \city{Pittsburgh, Pennsylvania}
    \country{USA}
}

\author{Xin Yi }
\authornote{Corresponding author.}
\orcid{0000-0001-8041-7962}
\email{yixin@tsinghua.edu.cn}
\affiliation{
    \institution{Tsinghua University}
    \city{Beijing}
    \country{China}
}

\author{Robert Xiao}
\orcid{0000-0003-4306-8825}
\email{brx@cs.ubc.ca}
\affiliation{
    \institution{University of British Columbia}
    \city{Vancouver, British Columbia}
    \country{Canada}
}

\author{Xinyi Fu}
\orcid{0000-0001-6927-0111}
\email{fuxy@tsinghua.edu.cn}
\affiliation{
    \institution{Tsinghua University}
    \city{Beijing}
    \country{China}
}

\author{Hewu Li}
\orcid{0000-0002-6331-6542}
\email{lihewu@cernet.edu.cn}
\affiliation{
    \institution{Tsinghua University}
    \city{Beijing}
    \country{China}
}

\renewcommand{\shortauthors}{Wang et al.}

\begin{abstract}
We present Roomify, a spatially-grounded transformation system that generates themed virtual environments anchored to users' physical rooms while maintaining spatial structure and functional semantics. Current VR approaches face a fundamental trade-off: full immersion sacrifices spatial awareness, while passthrough solutions break presence. Roomify addresses this through spatially-grounded transformation—treating physical spaces as ``spatial containers'' that preserve key functional and geometric properties of furniture while enabling radical stylistic changes. Our pipeline combines in-situ 3D scene understanding, AI-driven spatial reasoning, and style-aware generation to create personalized virtual environments grounded in physical reality. We introduce a cross-reality authoring tool enabling fine-grained user control through MR editing and VR preview workflows. Two user studies validate our approach: one with 18 VR users demonstrates a 63\% improvement in presence over passthrough and 26\% over fully virtual baselines while maintaining spatial awareness; another with 8 design professionals confirms the system's creative expressiveness (scene quality: 5.95/7; creativity support: 6.08/7) and professional workflow value across diverse environments.
\end{abstract}





\maketitle

\begin{figure*}[t]
    \centering
    \includegraphics[width=\textwidth]{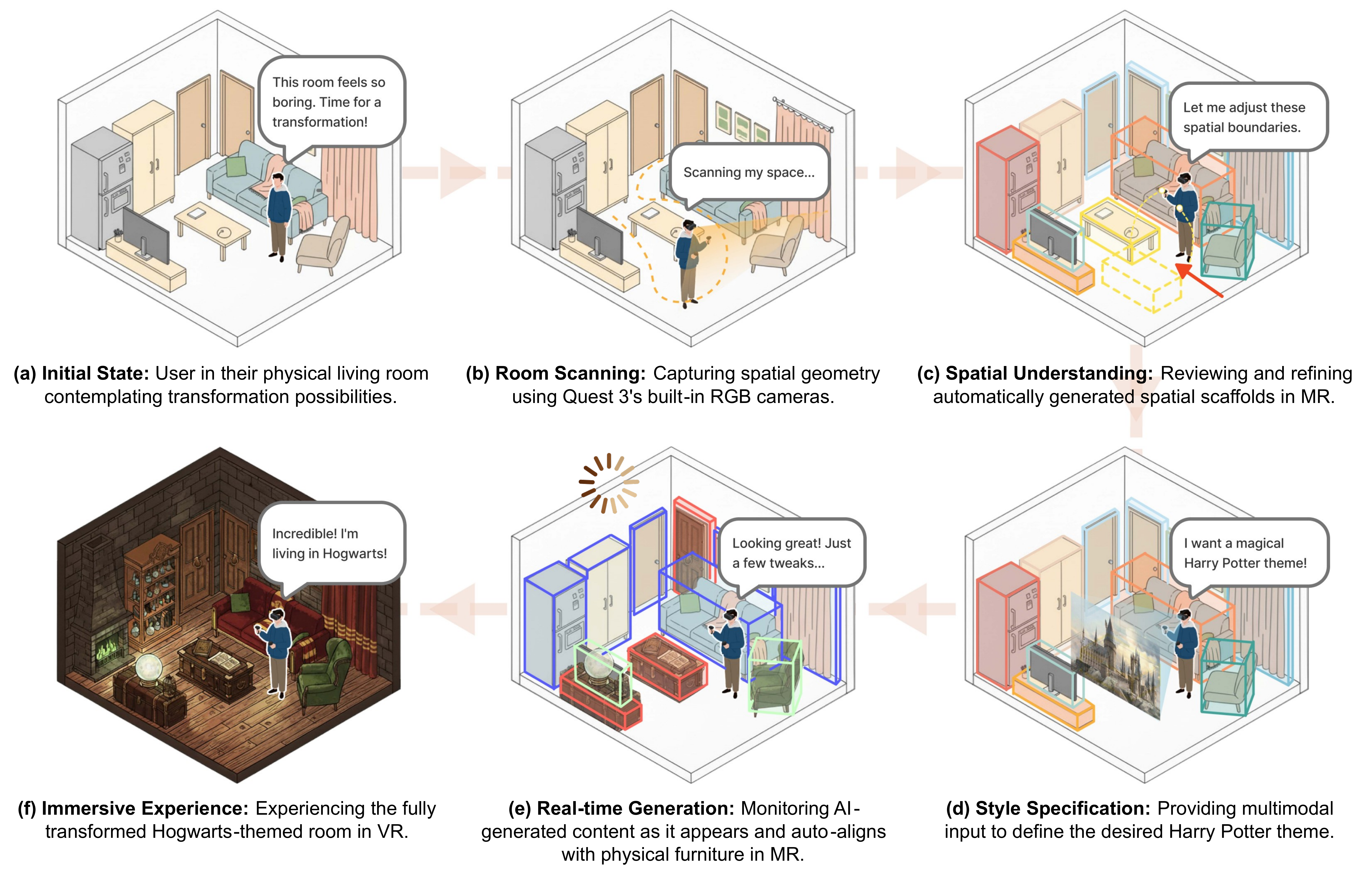}
    \caption{\textbf{The Roomify user journey from physical room to themed virtual environment.} Users begin in their real space (a), scan the room geometry (b), review spatial understanding results (c), specify their desired style through multimodal input (d), observe and adjust real-time generation in MR (e), and finally immerse themselves in the transformed environment in VR (f). The system preserves spatial layout and functional semantics of furniture while enabling radical stylistic transformation, allowing users to inhabit fantastical worlds grounded in their familiar physical space.}
    \label{fig:teaser}
\end{figure*}

\section{Introduction}

Virtual reality headsets are increasingly adopted in home environments for immersive entertainment and creative professional practices. However, this immersion comes at a fundamental cost: VR systems typically isolate users from their physical surroundings, creating spatial disorientation and safety risks \cite{das2024exploring, von2025qualitative, jelonek2023vrtoer}. Empirical studies demonstrate that users frequently breach safety boundaries even when visible warnings are present \cite{tseng2024understanding}, while research reveals that incorporating physical correspondence into VR experiences significantly improves both user experience and task performance \cite{cheng2023interactionadapt, budhiraja2015s, simeone2015substitutional}. For creative professionals, spatially-grounded immersive environments enhance understanding of spatial qualities like size, scale, and circulation \cite{wu2025two, strand2020virtual, chang2022influences}, with designers benefiting from MR-based on-site visualization and in-situ authoring tools \cite{dan2021holodesigner, lee2025imaginatear}.

These phenomena highlight the significant role of \emph{spatial awareness} in VR experiences: users must continuously know where their bodies and surrounding objects are while acting in a virtual scene. Studies confirm that spatial awareness is influenced by geometric features, semantic information, and spatial layouts---and when these sources are integrated, awareness is significantly enhanced \cite{Draschkow2017, Baltaretu2024, Alam2025}. For home entertainment, maintaining spatial awareness becomes inseparable from safety: virtual experiences must render furniture and boundaries legible as part of the spatial experience rather than as sporadic warnings \cite{chen2023virtualsafety}.

Current approaches to bridging physical and virtual worlds face a fundamental design dilemma: a trade-off between making spatial structure continuously perceptible and preserving uninterrupted presence \cite{chen2023virtualsafety}. Traditional boundary systems like Guardian provide only 2D perimeter warnings \cite{wu2023investigating}, failing to represent actual room geometry or furniture placement. Video passthrough solutions enable spatial awareness \cite{xiong2024petpresence, wang2022realitylens, hartmann2019realitycheck} but break immersion by constantly reminding users they are wearing a headset. Scene proxy methods that represent physical objects with virtual counterparts \cite{lin2020architect, sra2017oasis, simeone2015substitutional, yang2024dreamspace} maintain visual immersion but struggle with flexibility---either transforming only walkable areas \cite{sra2017oasis, cheng2019vroamer}, retrieving from limited asset libraries \cite{simeone2015substitutional, jiang2024situ, ipsita2024design}, or applying surface stylization without geometric freedom \cite{yang2024dreamspace, song2023roomdreamer}.

We present Roomify, a spatially-grounded transformation system that addresses this trade-off by achieving balance across \textbf{style diversity and consistency, spatial alignment, functional consistency, and user editability}. Our approach generates themed virtual environments anchored to users’ physical rooms while maintaining spatial structure and functional semantics of furniture. Rather than claiming to preserve full physical affordances, Roomify maintains each object’s approximate volume, footprint, and major contact surfaces while keeping its high-level role (e.g., seating, storage, temperature regulation) consistent with the physical room. This combination supports spatial awareness and gross locomotion and body-support actions (e.g., walking around obstacles, sitting, leaning). 

The system implements an end-to-end pipeline integrating room scanning \cite{liu2025slam3r}, AI-driven spatial understanding \cite{SpatialLM}, style-aware prompt generation, and coordinated image/3D generation with intelligent scene assembly. To address registration errors while providing fine-grained creative control, we developed a cross-reality authoring tool leveraging complementary MR and VR modalities \cite{von2025qualitative, auda2023scoping}. Users manipulate wireframe scaffolds and provide multimodal style specifications in MR mode for spatially-grounded refinement, then seamlessly transition to VR mode for immersive preview. This dual-mode approach enables creators to maintain both spatial constraints and creative vision throughout the authoring process.

Our evaluation through two user studies with 18 general VR users and 8 design professionals validates the system across diverse contexts---from themed entertainment (gaming, immersive movie viewing) to creative prototyping (interior design exploration, client visualization, media storyboarding). Roomify achieves 63\% improvement in presence over passthrough and 26\% over fully virtual baselines, while maintaining significantly better spatial awareness than fully virtual approaches. Design professionals rated scene quality at 5.95/7 compared to 4.50 for text-to-3D and 3.41 for AI re-texturing, with creativity support scores of 6.08/7.

This work makes three primary contributions:

\begin{itemize}
    \item \textbf{A spatially-grounded generation pipeline} that balances stylistic flexibility with geometric preservation and functional consistency, enabling personalized room transformation while maintaining spatial alignment through AI-driven spatial understanding and style-aware 3D generation.    
    \item \textbf{A cross-reality authoring tool} combining MR's spatial grounding with VR's immersive preview, providing intuitive controls for iterative refinement of room-scale transformation workflows.
    \item \textbf{Comprehensive empirical validation} through two user studies examining immersion, spatial awareness, and creative expressiveness, demonstrating significant improvements in presence while maintaining spatial alignment compared to existing approaches.
\end{itemize}

\section{Related Work}

\subsection{Integrating Physical and Virtual Environments}

Virtual reality systems face a fundamental tension between achieving deep immersion and maintaining spatial awareness in physical environments \cite{das2024exploring, von2025qualitative}. This challenge has driven researchers to explore various integration strategies.

\textbf{Boundary and Passthrough Systems.} Commercial systems like Oculus Guardian display grid overlays when users approach play area limits \cite{wu2023investigating}. However, users frequently breach boundaries during rapid movements, and these non-diegetic overlays disrupt immersion \cite{tseng2024understanding, hartmann2019realitycheck}. Video passthrough solutions enable spatial awareness and object manipulation \cite{xiong2024petpresence, wang2022realitylens, hartmann2019realitycheck, mcgill2015dose, das2024exploring, guo2023synchronous} but fundamentally break immersion—a ``dream collapsing'' effect \cite{knibbe2018dream}. Selective approaches like RealityCheck \cite{hartmann2019realitycheck} and RealityLens \cite{wang2022realitylens} blend physical objects into virtual scenes, but treat physical and virtual elements as separate layers rather than unified experiences.

Roomify establishes the physical room as the spatial foundation for transformation. Unlike boundary systems that interrupt immersion or passthrough methods that fragment experiences, Roomify integrates spatial structure directly into the generative process, creating personalized transformations while preserving spatial and semantic logic.

\textbf{Substitutional and Proxy-Based Approaches.} Scene proxy methods represent physical objects with virtual counterparts to maintain visual immersion \cite{lin2020architect, sra2017oasis, jiang2024situ, ipsita2024design, simeone2015substitutional, yang2024dreamspace, cheng2019vroamer}. However, these systems face flexibility and personalization limitations: some transform only walkable areas while ignoring furniture \cite{sra2017oasis, cheng2019vroamer}, others retrieve models from limited asset libraries \cite{simeone2015substitutional, jiang2024situ, ipsita2024design}, and still others apply surface stylization without geometric transformation freedom \cite{yang2024dreamspace, song2023roomdreamer}. Systems like GradualReality \cite{seo2024gradualreality} preserve tactile feedback through physical proxies but struggle when physical layouts don't correspond to virtual contexts.

Two systems warrant detailed comparison. Reality Skins \cite{shapira2016reality} reconstructs rooms with depth sensing and optimizes placement of pre-authored assets to maximize tactile alignment in designer-authored worlds. In contrast, Roomify focuses on generative, theme-driven transformation using LLM- and diffusion-based generation on 3D scaffolds, targeting end-user authoring workflows rather than constrained optimization over fixed asset libraries. TransforMR \cite{kari2021transformr} performs real-time substitution of moving agents (e.g., pedestrians) in outdoor handheld video-see-through MR, using monocular RGB to segment and overlay preset virtual characters. Roomify instead targets room-scale indoor VR/MR experiences, building complete 3D semantic scaffolds and applying scene-wide thematic transformations to walls, furniture, and decor.

\subsection{3D Scene Understanding and Generative Stylization}

\textbf{Spatial Understanding.} The evolution from point cloud processing to scene semantic understanding has enabled VR/MR systems to interpret physical environments as structured spaces \cite{reiners2021combination}. Modern pipelines establish hierarchical representations using datasets like ScanNet \cite{dai2017scannet} and structured graphs \cite{armeni20193d, wu2021scenegraphfusion} encoding spatial relationships and functional dependencies \cite{zhang2025open}. Neural SLAM advances like SLAM3R \cite{liu2025slam3r} and MASt3R-SLAM \cite{murai2025mast3r} enable real-time reconstruction, while systems like SpatialLM \cite{SpatialLM} and SceneScript \cite{avetisyan2024scenescript} generate structured scene descriptions through large language models. However, traditional pipelines terminate at analysis rather than enabling transformation.

\textbf{Text-to-3D and Scene Generation.} Text-to-3D methods like DreamFusion \cite{poole2022dreamfusion} and Magic3D \cite{lin2023magic3d} enable content creation from natural language, while room-scale systems like Text2Room \cite{hollein2023text2room} generate entire environments. However, these struggle with spatial consistency and constraint preservation. Constraint-aware systems like ControlRoom3D \cite{schult2024controlroom3d} improve layout control but offer limited style customization.

\textbf{Scene Stylization and Editing.} Recent vision and graphics work provides powerful building blocks for controllable stylization. ControlNet \cite{zhang2023adding} adds spatial conditioning to diffusion models for structure-preserving image stylization. InstructNeRF2NeRF \cite{haque2023instruct}, InstructGS2GS \cite{igs2gs}, and NeRF-Art \cite{wang2023nerf} edit existing NeRF/3DGS scenes using text instructions, iteratively updating representations for text-guided appearance changes with cross-view consistency. Styl3R \cite{wang2025styl3r} predicts stylized 3D Gaussians from sparse views and style images, while ArtiScene \cite{gu2025artiscene} assembles artistic 3D scenes from text via 2D layout intermediaries. Style transfer approaches like StyleMesh \cite{hollein2022stylemesh} and DreamSpace \cite{yang2024dreamspace} modify surface appearance but without geometric transformation freedom. Gaussian/volume-based stylization methods \cite{xu2025ssgaussian, huang2022stylizednerf} target scene-level style but can degrade object-level affordances.


Roomify does not introduce new 3D stylization algorithms; instead, it orchestrates existing 2D/3D generative tools atop semantic reconstruction of real home environments for theme-driven, room-scale transformations aligned with physical layout and functional roles. Unlike conventional scene understanding systems that conclude with static descriptions, our approach employs spatial models as dynamic containers enabling radical stylistic transformation while preserving spatial logic.

\subsection{AI-Assisted Spatial Authoring}

The democratization of VR content creation has shifted toward accessible, AI-assisted authoring systems \cite{huang2025personalized}. Recent systems enable non-expert creation through intuitive interactions: LLMR \cite{de2024llmr} and LLMER \cite{chen2025llmer} demonstrate natural language scene editing, while VRCoPilot \cite{zhang2024vrcopilot} introduces mixed-initiative authoring where users provide layout sketches and AI generates arrangements within constraints. MineVRA \cite{santarnecchi2025minevra} extends this through context-aware generation based on narrative contexts.

Critical challenges involve maintaining user agency while maximizing automation benefits. EchoLadder \cite{hou2025echoladder} decomposes AI modifications into transparent suggestions, while research reveals user expectations for context awareness, edit memory, and spatial reasoning in generative AI interfaces \cite{aghel2024people}. DreamCrafter \cite{vachha2025dreamcrafter} combines direct manipulation with generative backends; its MagicCamera feature enables furniture-level style transfer and scene generation. Roomify extends beyond furniture-level transformation to include structural and contextual elements (walls, floors, skyboxes), and grounds generation in actual physical room structure rather than arbitrary virtual scenes.

Complementary developments in computational design provide foundations for spatial transformation. Systems like Interactive Interior Design Recommendation \cite{zhang2023interactive} generate concepts from text and images, while C2Ideas \cite{hou2024c2ideas} automates color scheme generation. However, these typically operate in 2D or focus on isolated aspects without ensuring holistic spatial coherence.

Unlike text-to-scene methods \cite{zhang2024vrcopilot, hollein2023text2room} that generate environments from scratch, Roomify uses spatial understanding results as scaffolds, employing the real room as canvas for virtual environment creation. This provides enhanced automation eliminating manual object creation, and reality grounding leveraging users' actual room structure for intuitive spatial navigation. Our cross-reality workflow combines MR spatial grounding with VR immersive preview, enabling users to maintain both spatial constraints and creativ   e vision throughout the authoring process.

\section{Formative Study}

To establish evidence-based design principles for spatially-grounded virtual environment transformation, we conducted a formative study combining interviews with experienced VR users and professional designers engaged in VR-assisted spatial design. This early-stage research investigated needs, expectations, and workflows when integrating physical and virtual spaces.

\subsection{Methodology}

We recruited eight participants (Tab.~\ref{tab:formative_participants}) to capture perspectives on both everyday VR use and professional design workflows. Four were experienced VR users (P1--P4) with substantial VR experience (M = 50.6 hours), including active gamers familiar with titles such as \textit{Half-Life: Alyx} and \textit{Beat Saber}. The remaining four (P5--P8) were professional spatial or interaction designers who regularly employ VR/MR in practice (e.g., product, exhibition, and experience design).

We conducted 90-minute semi-structured interviews organized around three themes: (1) physical-virtual relationship conceptualization, including preferences for object preservation versus transformation; (2) control preferences regarding automation versus manual control; and (3) style consistency and immersion factors, including aesthetic coherence and spatial alignment. 

\begin{table*}[t]
\centering
\caption{Participant Demographics and VR Experience in Formative Study.}
\label{tab:formative_participants}
\small
\renewcommand{\arraystretch}{1.2}
\setlength{\tabcolsep}{4pt}
\begin{tabular}{@{}p{0.03\linewidth} p{0.05\linewidth} p{0.05\linewidth} p{0.12\linewidth} p{0.18\linewidth} p{0.18\linewidth} p{0.28\linewidth}@{}}
\toprule
\textbf{ID} & \textbf{Age} & \textbf{Gender} & \textbf{VR Exp. (hrs)} & \textbf{HMDs Used} & \textbf{User Group} & \textbf{Desired Immersion Contexts}\\
\midrule
P1 & 25-34 & Female & 10-20 & PS VR & General User & Gaming\\
P2 & 25-34 & Female & 20-50 & Meta Quest, PICO series & General User & Immersive creative workspace\\
P3 & 25-34 & Female & 50-100 & HTC Vive, PS VR & General User & Immersive spatial design showcase\\
P4 & 18-24 & Male & 20-50 & PICO Series & General User & Gaming\\
P5 & 25-34 & Female & 50-100 & PS VR & Expert Designer & Immersive creative workspace\\
P6 & 18-24 & Female & 100+ & Meta Quest & Expert Designer & Immersive spatial design showcase\\
P7 & 18-24 & Female & 20-50 & PICO Series & Expert Designer & Gaming\\
P8 & 18-24 & Female & 20-50 & PICO Series & Expert Designer & Gaming\\
\bottomrule
\end{tabular}
\end{table*}

\subsection{Findings}

Analysis revealed four consistent themes guiding system design:

\textit{Intent-Driven Style Consistency.} Participants welcomed environmental transformation but rejected arbitrary style variations disrupting narrative coherence. P1 emphasized emotional resonance between environment and activity, while P4 preferred ``real-yet-beyond-real'' transformations preserving function while creatively altering form. Design professionals stressed that narrative coherence between transformed environments and physical sites is critical---mismatched styles undermine users' sense of place. This informed Roomify's scene-level style conditioning guided by user intent and context.

\textit{Functional Consistency Over Geometric Fidelity.} Participants prioritized maintaining object function and spatial navigation over exact geometric appearance. P4 articulated this as preferring forms that ``bend without breaking function,'' while P5 noted that meaning arises from physical engagement rather than abstract calculation. P6 emphasized that real-world scale serves as an essential reference frame for judging feasibility. These findings guide Roomify's approach of maintaining functional semantics and spatial relationships while enabling radical aesthetic transformation.

\textit{Hierarchical Control Preferences.} Participants expressed nuanced preferences for AI-human collaboration. P3 proposed selective user control where ``users lock key items while AI stylizes the rest,'' while P8 desired maximum automation with fine-tuning capabilities. Designers valued how this synergy amplifies inspiration while allowing direct artistic expression, preferring to remain within the headset for iterative testing while the system handles background generation. This informed Roomify's three-tier control architecture: global style specification, object-level semantic management, and fine-grained attribute adjustments.

\textit{Context-Sensitive Spatial Requirements.} Spatial preservation needs vary by use context. Gaming applications prioritized immersive transformation with navigational safety, while creative workspaces emphasized functional object preservation. P6 highlighted that immersive environments facilitate rapid iteration, while P5 envisioned democratizing design by enabling stakeholders to intuitively evaluate proposals within authentic spatial contexts. These requirements reflect two use families---exploratory entertainment versus productive work---each demanding different balances between transformation and spatial stability.

\subsection{Design Requirements}

Based on our analysis, we established four design principles guiding Roomify's implementation:

\begin{itemize}
    \item \textbf{Style Diversity and Consistency.} Support coherent stylistic transformation with localized customization based on user intent. Thematic transformations should reinforce rather than distract from narratives staged in the physically grounded space.
    
    \item \textbf{Spatial Alignment.} Preserve essential geometric relationships, navigation paths, and stable correspondence between virtual elements and physical room geometry to maintain user confidence and spatial logic.

    \item \textbf{Functional Consistency and Geometric Preservation.} Maintain each object’s high-level function (e.g., seating, storage, temperature regulation) while preserving its approximate bounding box, footprint, and major contact surfaces throughout stylistic transformation, to facilitate key everyday affordances such as sit-ability, leaning, and obstacle avoidance.

    \item \textbf{User Editability.} Provide hierarchical control combining AI automation with selective human oversight, available directly within immersive MR/VR workflows for in-situ iteration.
\end{itemize}

\section{Spatially-Grounded Scene Generation Pipeline}

\subsection{Pipeline Overview}

\begin{figure*}[t]
    \centering
    \includegraphics[width=\textwidth]{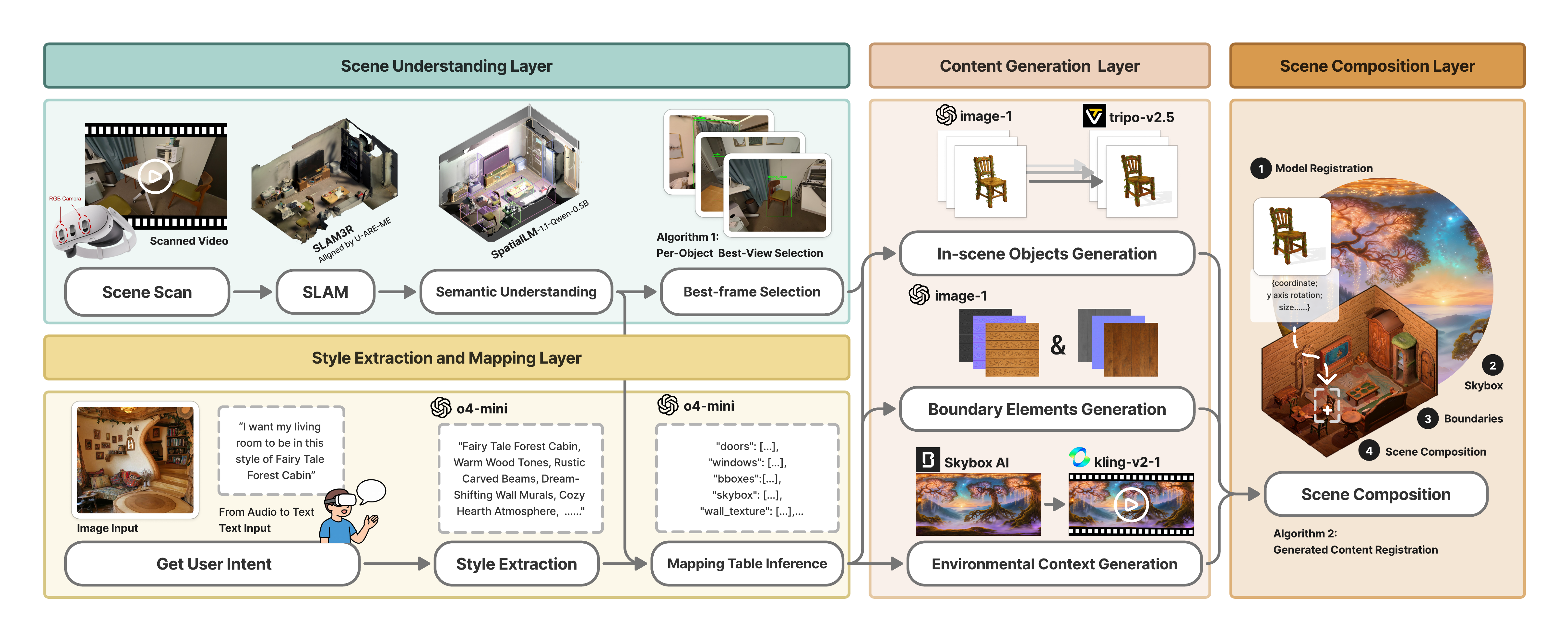}
    \caption{Spatially-grounded scene generation pipeline showing the four-stage transformation process from physical room capture to stylized virtual environment.}
    \label{fig:pipeline_overview}
\end{figure*}

Building on the design requirements from our formative study, Roomify implements a four-stage transformation pipeline that converts physical environments into personalized virtual spaces while preserving spatial layouts and functional semantics (Fig.~\ref{fig:pipeline_overview}). The system treats physical rooms as spatial containers, maintaining spatial awareness during immersive transformation.

The pipeline accepts two inputs: a 30-60 second monocular RGB video captured with Meta Quest 3, and multimodal user prompts (text and/or reference images) specifying aesthetic preferences. Output comprises a complete stylized virtual environment with structural elements (walls, floors), environmental context (skybox), and transformed objects that maintain semantic roles while adopting target aesthetics.

The four stages are: \textit{Scene Understanding} (geometric reconstruction and spatial parsing), \textit{Style Extraction and Mapping} (coherent specifications and transformation rules), \textit{Content Generation} (stylized assets), and \textit{Scene Composition} (spatial registration into a cohesive environment). We extend JSON-based scene representation approaches \cite{chen2025llmer}, treating each component as a spatial container that preserves semantic logic while enabling aesthetic transformation.

\subsection{Spatial Scene Understanding}

The understanding stage establishes geometric and spatial foundations for structure-preserving transformation, addressing the \textit{spatial alignment} and \textit{functional consistency} requirements.

Geometric reconstruction uses SLAM3R \cite{liu2025slam3r} to process monocular RGB video, generating dense point clouds with frame-wise camera poses. Spatial alignment follows using U-ARE-ME \cite{patwardhan2024uaremeuncertaintyawarerotationestimation} to estimate Manhattan axes and gravity direction, ensuring compatibility with structured indoor modeling systems \cite{dai2017scannet}.

The aligned point cloud undergoes spatial parsing by SpatialLM \cite{SpatialLM}, which identifies architectural elements (walls, doors, windows) and furniture objects, outputting structured descriptions with oriented bounding boxes encoding geometric properties and spatial categories. These outputs are serialized into a global scene JSON file. Each 3D bounding box functions as a \textbf{spatial scaffold} encoding geometric constraints and semantic information, providing the structural foundation for stylization while ensuring generated content maintains spatial relationships.

\subsection{Style Extraction and Mapping}

\begin{figure*}[t]
    \centering
    \includegraphics[width=\textwidth]{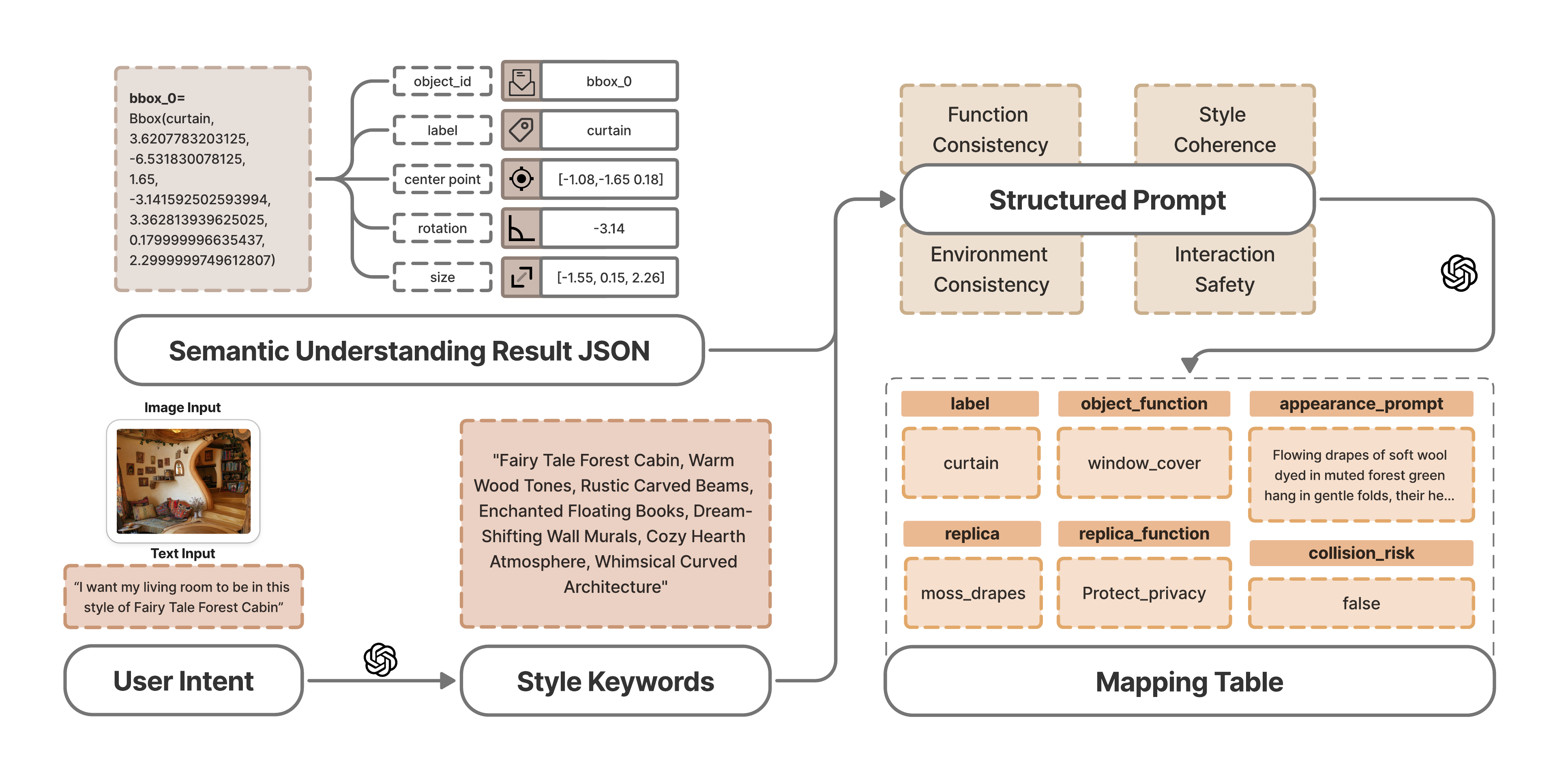}
    \caption{Style extraction and mapping workflow showing transformation from user intent and spatial understanding results to structured generation specifications. The process extracts style keywords from multimodal input and creates object-level mapping tables satisfying four criteria: function consistency, style coherence, environmental consistency, and interaction safety.}
    \label{fig:inference_overview}
\end{figure*}

The style extraction and mapping workflow (Fig.~\ref{fig:inference_overview}) transforms user intent into actionable generation specifications, implementing the \textit{style diversity and consistency} requirement by establishing global aesthetic constraints for object-level transformations.

\begin{figure*}[t]
    \centering
    \includegraphics[width=0.95\textwidth]{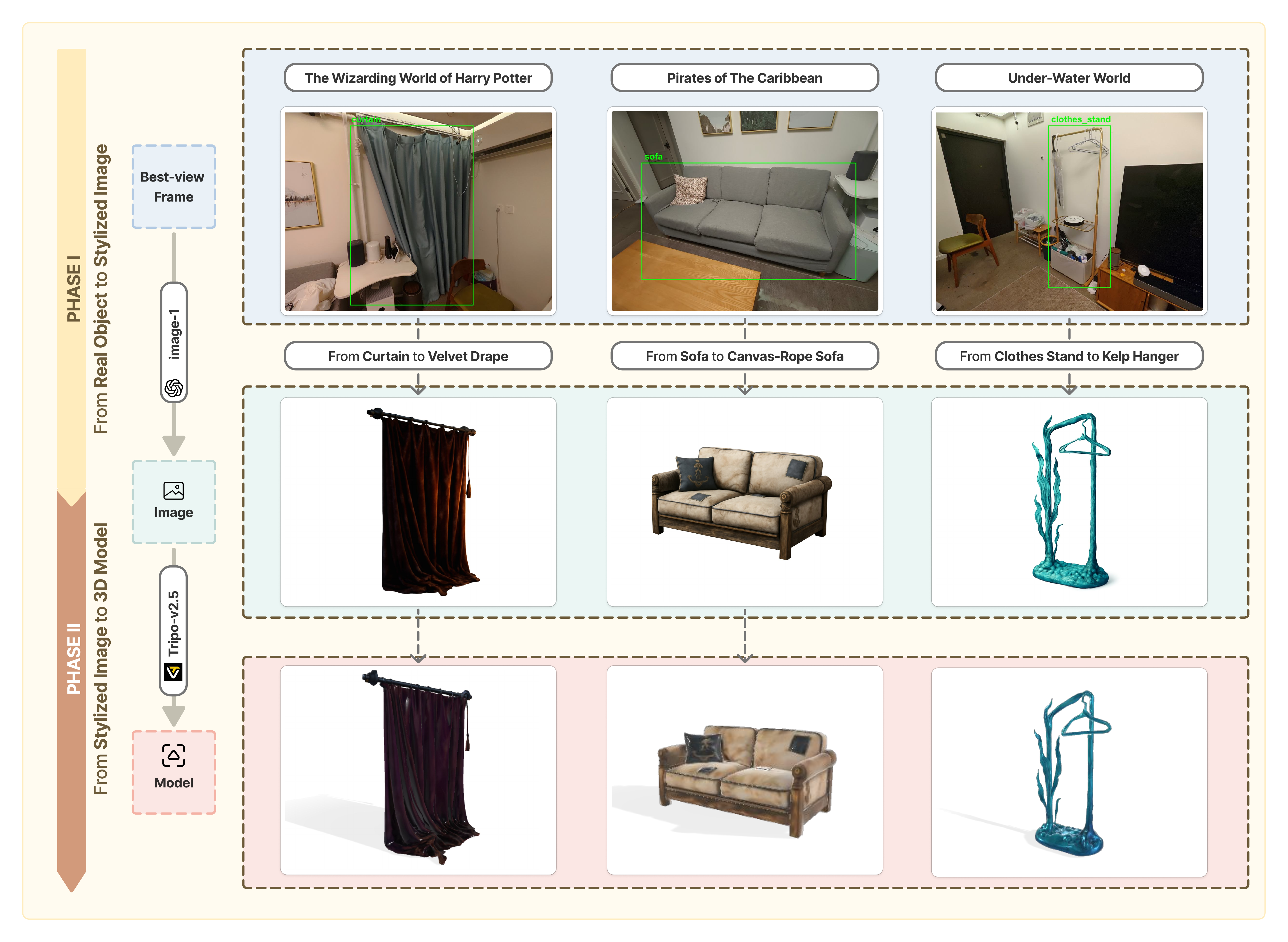}
    \caption{Reference-guided object generation workflow: best-view frame selection captures optimal geometry, style-aware image generation preserves spatial characteristics, and 3D model conversion completes the transformation. Examples show diverse styles while maintaining semantic recognition and spatial consistency.}
    \label{fig:Text-image-model}
\end{figure*}

Style specification begins with multimodal intent processing using a language model agent (o4-mini). The agent derives 4-8 structured style keywords encompassing style category, color palette, material properties, and atmospheric qualities, providing unified guidance for object-level generation and preventing style drift.

The extracted keywords and spatial scene representation are processed by a mapping agent generating a \textbf{transformation specification table} organizing components into three categories: (1) \textit{In-scene objects} (furniture, doors, windows); (2) \textit{Boundary elements} (walls, floors); (3) \textit{Environmental context} (skybox).

We operationalize design requirements into four objectives: (1) Functional Consistency---preserve object functional roles and task substitutability (e.g., seating remains recognizable as seating); (2) Style Coherence---ensure objects align with global style keywords; (3) Environmental Consistency---maintain harmony across inter-object relations while preserving approximate correspondence with real environment layout and scale; (4) Interaction Safety---infer potential collision risks for subsequent user adjustment.

For each object, the mapping agent analyzes spatial labels to infer \texttt{object\_function}, proposes a semantically compatible \texttt{replica}, and generates \texttt{appearance\_prompt} specifications plus \texttt{collision\_risk} assessment. Boundary mapping generates \texttt{texture\_prompt} specifications for seamless, tileable materials, while skybox generation receives \texttt{style\_prompt} and \texttt{negative\_constraint} specifications.

\subsection{Multi-Modal Content Generation}

The content generation stage transforms mapping specifications into visual assets through specialized pipelines for each component category. All generation processes execute in parallel, significantly reducing perceived waiting time.

\textbf{In-Scene Object Generation.} Object transformation employs a reference-guided three-step workflow (Fig.~\ref{fig:Text-image-model}). Best-view frame selection (Algorithm~1, Appendix) analyzes SLAM video to identify optimal viewing angles based on visibility, centering, and occlusion criteria, ensuring generated objects maintain geometric properties and orientational consistency with original positions.

GPT Image (\texttt{gpt-image-1}) processes prompts with selected best-view frames, generating stylized images preserving essential geometric properties. This intermediate image generation makes the process aware of object geometry while preserving orientation consistency for accurate registration. Tripo AI (v2.5) then converts images to lightweight 3D models optimized for real-time rendering. 

\textbf{Boundary Element Generation.} Wall and floor textures use PBR material generation. GPT Image generates seamless, tileable RGB textures with additional processing for metallic and normal maps enabling realistic lighting behavior.

\textbf{Environmental Background Generation.} Rather than preserving ceiling geometry, we generate dynamic skybox environments for enhanced spatial openness. The process uses Blockade Labs Skybox AI for static panoramic generation, followed by Kling-v2-1 for 10-second dynamic sequences with ambient audio, looped seamlessly through forward-reverse concatenation.

\begin{figure*}[t]
    \centering
    \includegraphics[width=1\textwidth]{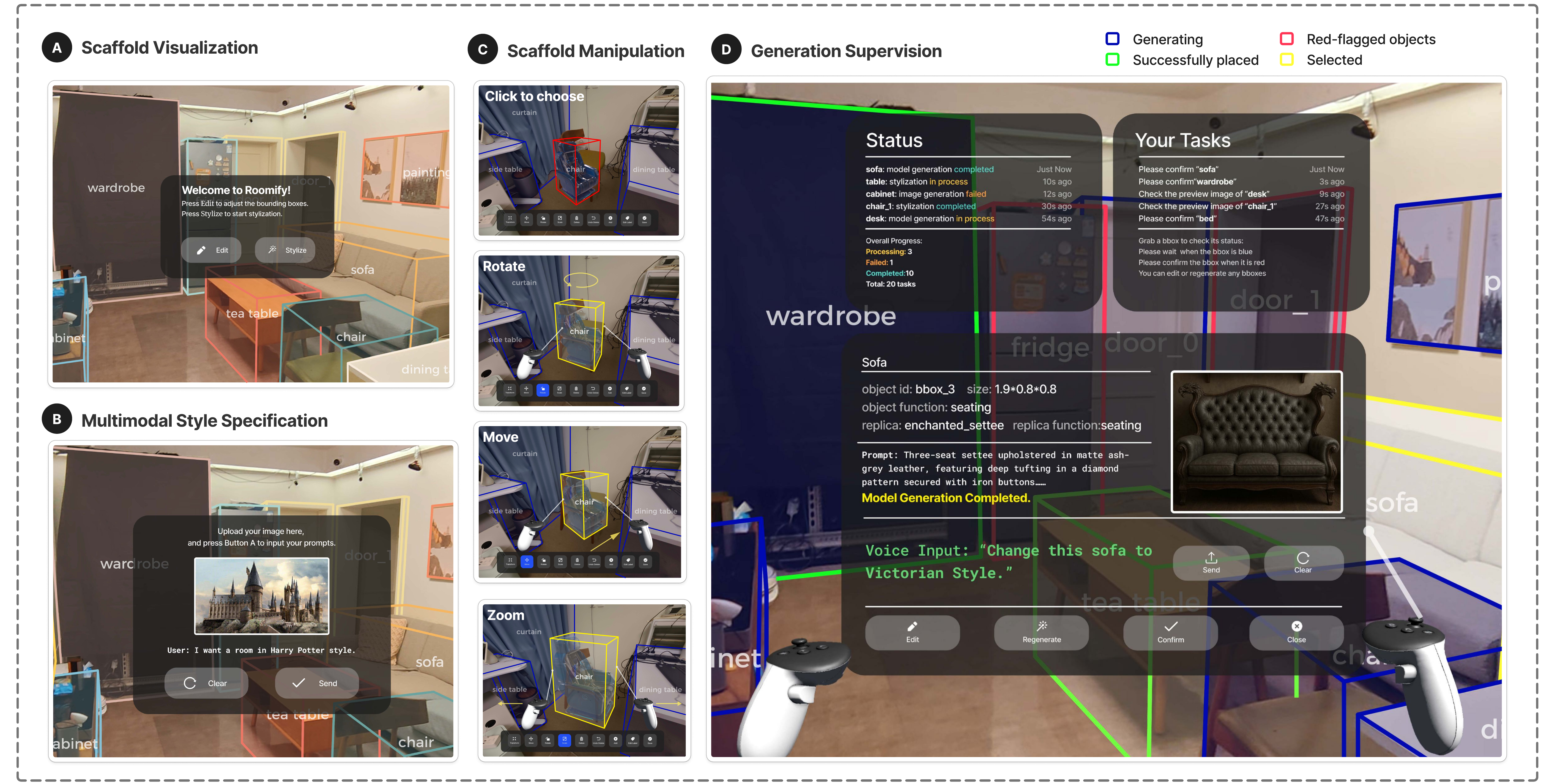}
    \caption{Cross-Reality Authoring Tool interface. (A) Spatial Scaffold Visualization displays detected objects with labeled, color-coded wireframe boundaries overlaid on the physical environment. (B) Multimodal Style Specification enables combining text descriptions and reference images. (C) Spatial Scaffold Manipulation provides controls for selection, rotation, translation, and scaling; selected objects become translucent to reveal underlying physical objects for accurate alignment. (D) Generation Process Supervision shows status panels with generation progress, object information, and voice-based refinement commands; wireframe colors indicate status (blue: generating, green: complete, red: requires attention).}
    \label{fig:authoring_overview}
\end{figure*}

\subsection{Scene Composition}

The final stage integrates generated assets through precise spatial registration maintaining geometric relationships with physical space, addressing the \textit{spatial alignment} requirement.

The Generated Content Registration Algorithm (Algorithm~2, Appendix) performs systematic integration using spatial scaffolds and best-view camera poses. Object registration begins with isotropic scaling to match scaffold dimensions, followed by orientation optimization using IoU maximization. Fine-grained scaling ensures objects remain within spatial boundaries while allowing reasonable geometric variation. Ground plane alignment ensures stable positioning.

Environmental integration applies boundary textures with 5cm outward offset for walls to prevent occlusion conflicts with doors and windows. Skybox integration provides global environmental context with unified lighting conditions.

\subsection{Implementation Details}

We implement the pipeline with a cloud-edge architecture. SLAM3R, U-ARE-ME, and SpatialLM are deployed on a GPU-backed server; monocular RGB video from Meta Quest 3 is uploaded and processed, returning JSON scaffolds. A laptop-based LangChain orchestrator invokes external generative APIs (o4-mini, gpt-image-1, Tripo AI, Blockade Labs Skybox, Kling-v2-1), serializing outputs into global scene JSON. A Flask service exposes REST endpoints for Unity-Python messaging. The Unity client loads generated assets at runtime via Meta Quest Link. Generation requests execute concurrently (2 for stylized-image, 9 for image-to-3D conversion).

\section{Cross-Reality Authoring Tool}

Building on the spatially-grounded generation pipeline, we present the Cross-Reality Authoring Tool---an interactive interface that transforms complex generative processes into intuitive creative workflows. The tool serves three functions: providing the primary interface for intent specification and content preview, enabling fine-grained control over generation results for spatial accuracy, and supporting seamless transitions between MR creation and VR experience modes.

The system implements a dual-modality design leveraging complementary strengths of Mixed Reality and Virtual Reality \cite{von2025qualitative, auda2023scoping}. MR mode utilizes the physical environment as a spatial scaffold for grounded editing and iterative refinement with direct reference to real-world constraints. VR mode enables immersive evaluation of transformed environments while maintaining spatial correspondence. This workflow addresses the \textit{user editability} requirement by balancing automated generation with precise human control. Recognizing that generative AI can produce inappropriate outputs, the tool incorporates multiple safeguards allowing users to detect and correct errors in real-time.

\subsection{Spatial Scaffold Manipulation} 

The authoring process establishes a manipulable digital representation of the physical environment. The system processes scene JSON from the backend and instantiates each detected spatial entity as an interactive 3D bounding box within the Unity-based frontend.

These bounding boxes serve dual functions: as \textbf{interactive handles} providing tangible proxies for spatial manipulation where stylized models inherit properties from parent boxes, and as \textbf{stateful containers} visualizing object properties and recording user modifications. The MR visualization renders each box with centered text labels indicating object categories and color-coded wireframes differentiating object semantics (Fig.~\ref{fig:authoring_overview}A).

For spatial calibration between the scanned scene and generated virtual content, we rely on Meta's Mixed Reality Utility Kit (MRUK) Spatial Anchors. During first use, users manually adjust the world origin position and rotation so that scaffolded walls and furniture align with real-world objects (typically 1--2 minutes). Once confirmed, our system creates a Spatial Anchor at this calibrated origin. On subsequent launches, the system resolves this anchor and automatically re-applies the stored transform, maintaining consistent spatial correspondence without recalibration.

\subsection{MR Mode: Integrated Creative Workspace}

MR mode provides comprehensive authoring through four interconnected modules (Fig.~\ref{fig:authoring_overview}) balancing automation with creative control.

\textbf{Multimodal Style Specification} (Fig.~\ref{fig:authoring_overview}B) captures creative vision through complementary modalities. Natural language input enables conceptual control through descriptive prompts with real-time transcription. Visual reference upload provides aesthetic guidance for qualities difficult to verbalize. The system synthesizes these into unified specifications, reducing ambiguity and the risk of outputs deviating from expectations.

\textbf{Spatial Scaffold Manipulation} (Fig.~\ref{fig:authoring_overview}C) enables precise geometric control. Users interact with bounding boxes through direct controller manipulation with visual feedback via yellow wireframe highlighting. Bimanual gestures control translation (controller average position), rotation (relative rotation), and scaling (controller distance). During manipulation, objects enter a translucent state revealing underlying physical objects for accurate alignment---enabling users to identify and correct inconsistencies from imperfect AI spatial reasoning. Users can also add or delete bounding boxes to address scan errors, with voice-based semantic editing for misclassification correction.

\textbf{Generation Process Supervision} (Fig.~\ref{fig:authoring_overview}D) transforms opaque AI generation into transparent workflows. Stateful wireframe visualization provides scene-wide status: blue indicates active generation, green confirms successful placement, and red flags objects requiring attention due to collision risks identified during Style Extraction and Mapping. Users must confirm red-flagged placements before VR mode entry. Individual object panels display generation metadata, AI-inferred prompts, and preview images. Voice adjustment instructions (e.g., ``Change this sofa to Victorian style'') trigger regeneration, enabling correction of unsatisfactory outputs without restarting the workflow.

\subsection{VR Mode: Immersive Experience}

VR mode completes the workflow by enabling full immersive experience. Transition triggers scene finalization including wall and floor textures, thematically consistent skybox environments, and contextual ambient audio for complete multi-sensory experiences.

The VR experience maintains bidirectional connectivity with MR mode for seamless transitions. This serves two functions: aesthetic iteration when experience quality requires adjustment, and dynamic adaptation to physical environment changes. If furniture is relocated, users can return to MR mode to adjust virtual representations, maintaining spatial correspondence throughout extended sessions.

\section{Study 1: Real-World User Experience Evaluation}

We conducted a controlled within-subjects study to evaluate Roomify's effectiveness in balancing immersive entertainment with spatial awareness in realistic home environments.

\textbf{RQ1:} Does spatially-grounded style transformation enhance entertainment experience and immersion compared to existing VR approaches?

\textbf{RQ2:} Can Roomify maintain spatial awareness during VR activities requiring physical movement?

\subsection{Experimental Design and Conditions}

\begin{figure*}[t]
\centering
\includegraphics[width=0.9\linewidth]{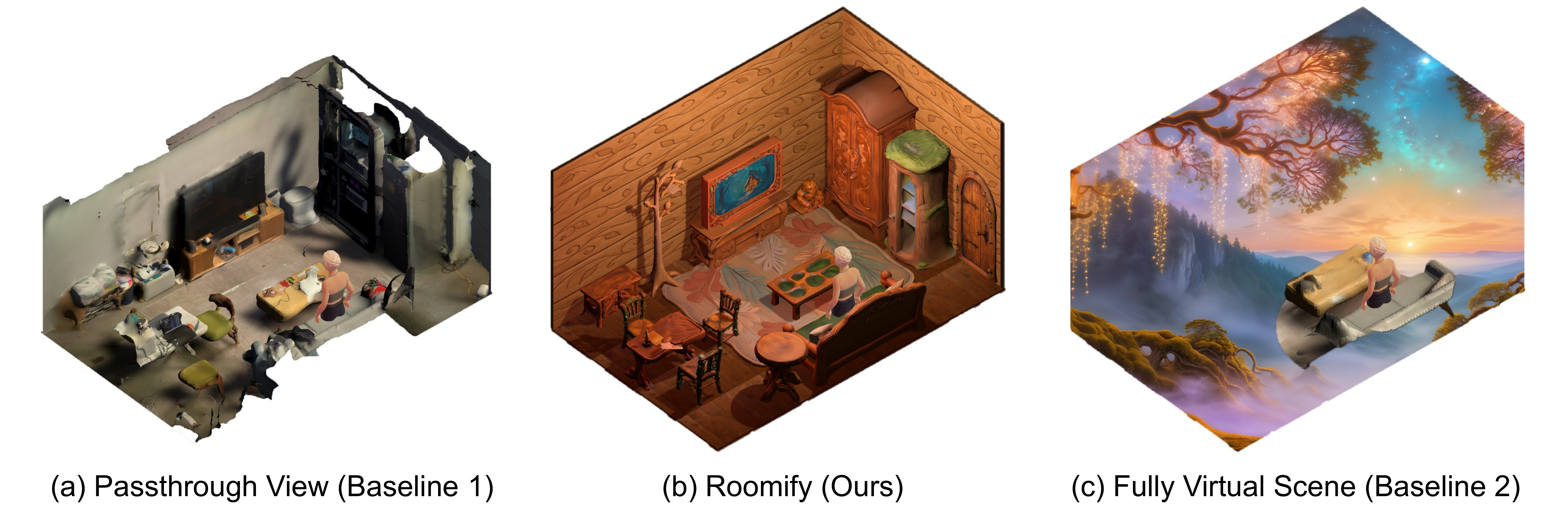}
\caption{Three VR environment conditions: (a) Passthrough baseline with direct real-world visibility, (b) Roomify's spatially-grounded transformation preserving spatial structure with thematic coherence, and (c) Fully virtual baseline with proximity-based boundary display.}
\label{fig:baslines}
\end{figure*}

We employed a within-subjects design comparing Roomify against two baseline conditions representing the current spectrum of commercial VR approaches (Fig.~\ref{fig:baslines}).

\begin{itemize}
    \item \textbf{Passthrough View} (Baseline 1) provides maximum spatial awareness through unmodified camera feed, representing standard mixed reality modes in devices like Quest 3 and Apple Vision Pro. Immersion is provided through virtual screens and gameplay elements, while safety is maximized through direct visibility of the real environment.
    \item \textbf{Fully Virtual Scene} (Baseline 2) delivers maximum immersion through 360° panoramic environments generated using Skybox AI to match entertainment themes. To avoid misleading affordances, no additional 3D objects are placed in the scene. For safety during physical movement, we implemented proximity-based boundary warnings displaying surroundings when users approach real objects within 1 meter---the standard mechanism in current VR systems \cite{tseng2024understanding, visionOS}.
    \item \textbf{Roomify} (Experimental Condition) applies spatially-grounded transformation to create thematically coherent environments while maintaining spatial structure and semantic consistency of the physical room, balancing safety and immersion.
\end{itemize}

Our primary goal was to compare how well each condition balances immersion and spatial awareness. For the Fully Virtual condition, we chose a ``skybox cinema'' setup without virtual furniture---isolating the effect of having versus not having a mapping to physical room structure, rather than conflating comparison with differences in scene richness. Adding virtual furniture would introduce confounds because virtual object positions might not reflect real obstacle locations, creating safety risks during locomotion consistent with Meta's Mixed Reality Health and Safety Guidelines\footnote{https://developers.meta.com/horizon/design/mr-health-safety-guideline/}. Similarly, participants could not disable the boundary system: on consumer headsets, disabling the guardian requires developer options and is not recommended for general users who may forget to re-enable it before moving.

All conditions include both immersion and safety components, representing the fundamental trade-off between spatial awareness and immersive transformation. Roomify aims to optimize this balance through structured transformation.

\subsection{Participants and Protocol}

We recruited 18 participants (10 female, 8 male) aged 20-32 years (M=23.33, SD=3.03) with prior AR/VR/MR experience (average familiarity: 3.33/5). Experience distribution: 5 participants with 1-5 sessions, 7 with 5-10 sessions, 6 with 10-30 sessions. The study received IRB approval; participants provided informed consent and received \$20 compensation.

The study environment was a realistic living room (5.10m × 3.15m × 2.40m, as seen in Fig.~\ref{fig:baslines}a) containing three doors and approximately fifteen furniture pieces, providing authentic spatial complexity.

\textbf{Familiarization (10 min):} Participants explored the physical environment and were randomly assigned one of six entertainment scenarios: three games (\textit{Plants vs. Zombies}, \textit{Space Station Robot Shooter}, \textit{Fishing Master}) and three movies (\textit{Pirates of the Caribbean}, \textit{Jurassic World}, \textit{Harry Potter}).

\textbf{Environment Creation (20-25 min):} Participants learned the Roomify interface and created personalized environments. To ensure consistency and focus on transformation effectiveness, we provided standardized pre-recorded room scan video \cite{liu2025slam3r}, isolating style transformation evaluation from scanning variability.

\begin{figure}[t]
\centering
\includegraphics[width=\columnwidth]{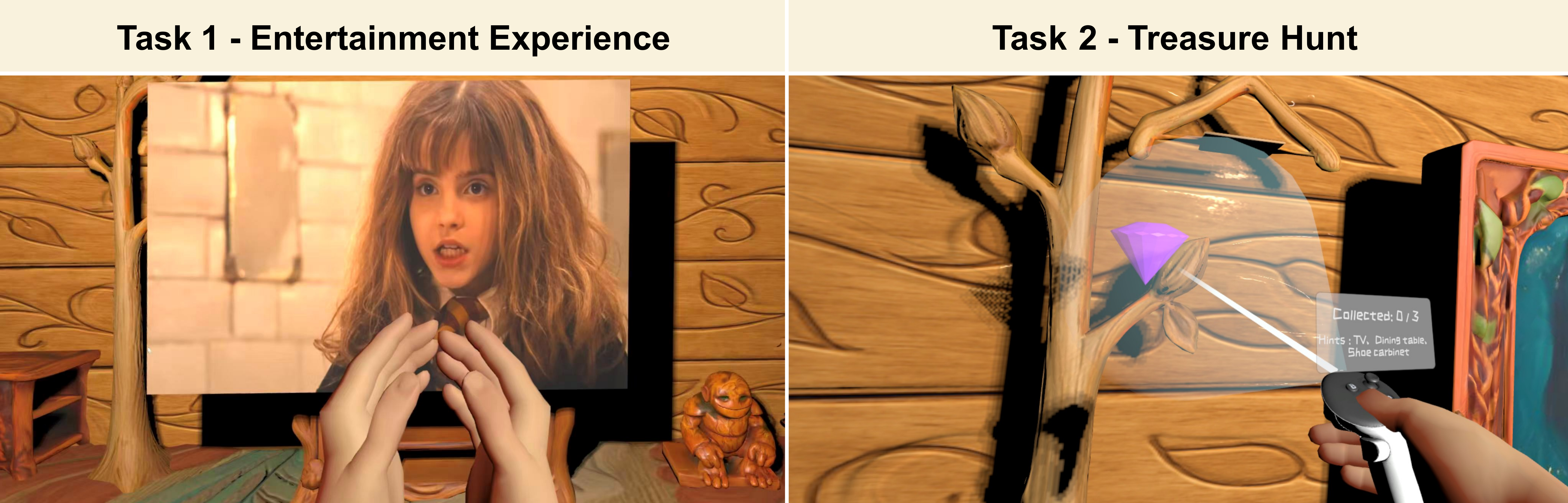}
\caption{Study 1 tasks: Task 1 entertainment experience with \textit{Harry Potter} content; Task 2 treasure hunt requiring spatial navigation and furniture interaction.}
\label{fig:tasks}
\end{figure}

\textbf{Task Execution (20-25 min):} Participants completed two tasks (Fig.~\ref{fig:tasks}) in each condition using Latin square counterbalancing:
\begin{itemize}
    \item \textit{Task 1 - Entertainment Experience:} Participants engaged with assigned content on a virtual screen for 3-5 minutes, assessing presence and engagement.
    \item \textit{Task 2 - Treasure Hunt:} Starting seated at a dining table, participants located three randomly positioned gems using a virtual flashlight, then navigated to and sat on the real sofa. This tested spatial navigation, collision avoidance, and orientation maintenance. Participants reported safety incidents while we recorded completion time.
\end{itemize}

\textbf{Assessment (15 min):} Questionnaires and interviews exploring experience, performance, and preferences.

The study protocol was reviewed and approved by the Institutional Review Board (IRB) of Tsinghua University.

\begin{figure*}[t]
\centering
\includegraphics[width=\linewidth]{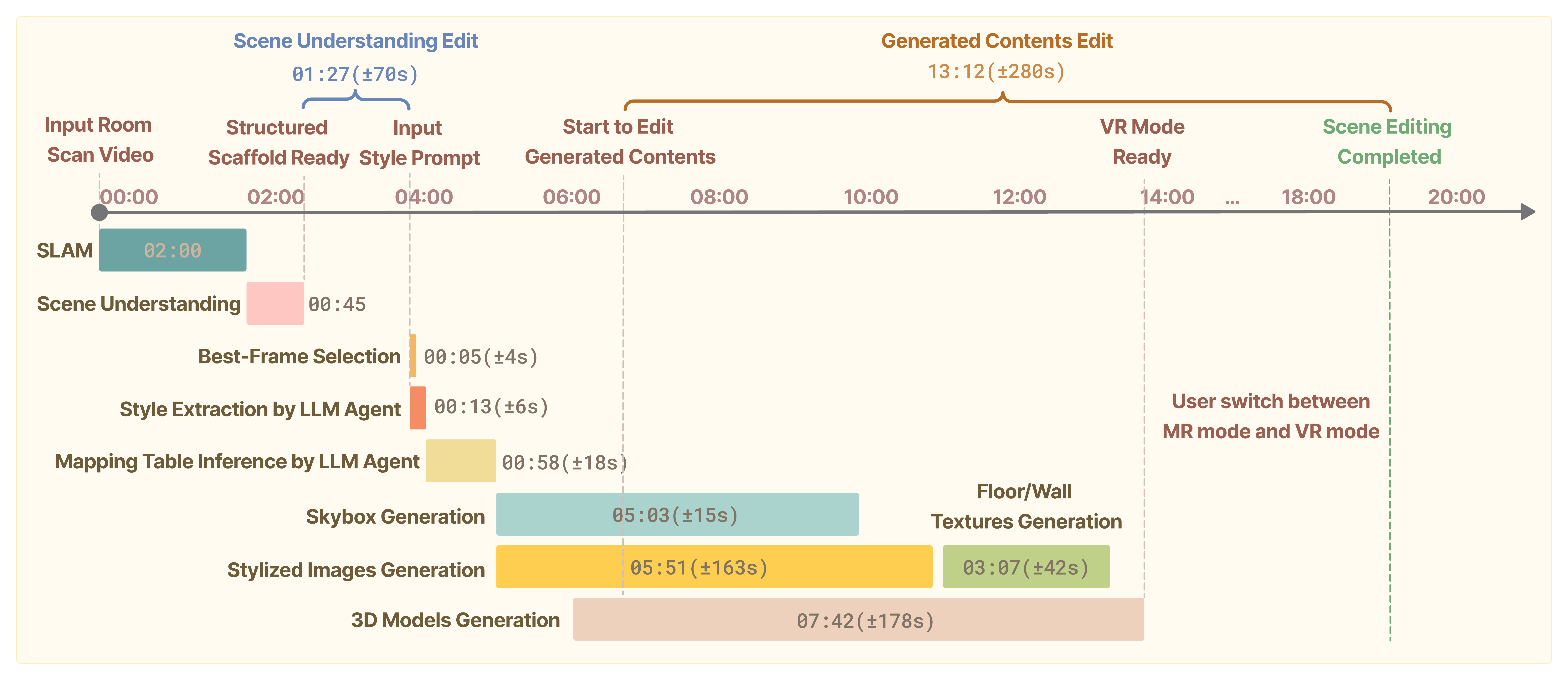}
\caption{Roomify creation pipeline timeline showing parallel processing stages. Users actively edit during content generation rather than waiting passively. Numbers indicate mean completion times with standard deviations.}
\label{fig:timeline}
\end{figure*}

\begin{table*}[t]
\centering
\caption{User operation frequencies during environment creation phases (mean ± standard deviation, N=18).}
\label{tab:editing-operations}
\begin{tabular}{lcccccccc}
\toprule
\textbf{Edit Phase} & \textbf{Duration} & \textbf{Move} & \textbf{Rotate} & \textbf{Scale} & \textbf{Delete} & \textbf{Edit Label/} \\
& & & & & & \textbf{Regenerate} \\
\midrule
Scene Understanding & 1:27 (±70s) & 0.29 (±0.47) & 1.14 (±1.56) & 0.14 (±0.36) & 0.14 (±0.36) & 0.07 (±0.27) \\
Generated Content & 13:12 (±280s) & 10.29 (±6.66) & 4.07 (±4.46) & 2.57 (±1.91) & 0.50 (±0.76) & 0.78 (±1.35) \\
\bottomrule
\end{tabular}
\end{table*}

\subsection{Measurements and Analysis}

\textbf{Subjective Measures:} System Usability Scale (SUS) \cite{brooke2013sus} for authoring tool effectiveness; Igroup Presence Questionnaire (IPQ \cite{schubert2001experience}, 4 items) for presence; User Experience Questionnaire-Short (UEQ-S \cite{hinderks2017design}, 10 items) for interaction quality; Spatial Awareness Scale (6 items) based on established instruments \cite{ring2024development}.

\textbf{Objective Performance:} Task 2 completion time and self-reported safety incidents; NASA-TLX \cite{hart2006nasa} for cognitive workload.

Statistical analysis employed mixed ANOVA with Aligned Rank Transform (ART) \cite{elkin2021aligned} for non-parametric factorial analysis, with Greenhouse-Geisser corrections where necessary. Post-hoc comparisons used Holm-corrected paired t-tests. Qualitative analysis applied thematic analysis \cite{mcdonald2019reliability} to interview transcripts.

\subsection{Results}

\subsubsection{Authoring Tool Performance}

Participants utilized Roomify's cross-reality workflow, transitioning between MR and VR modes to complete environment transformation through two phases: spatial understanding refinement and generated content adjustment.

Creation time analysis (Fig.~\ref{fig:timeline}) revealed efficient workflows. To compute operation durations, we combined timestamped interaction logs with screen-capture videos, annotating each session into labeled phases (style specification, spatial refinement, asset generation). Phase duration was defined as the time span from the first to the last event within that phase; because phases could overlap (e.g., participants repositioning objects while assets regenerated), durations were measured independently and may not sum to total session time.

Participants completed full environment transformations in 19 minutes 46 seconds on average (SD = 5.0 min), with parallel processing enabling continuous engagement. Spatial understanding outputs required minimal intervention (M = 1:27, SD = 70s), validating reconstruction accuracy. Generated content refinement occupied most creation time (M = 13:12, SD = 4:40), reflecting both generation latency and participants' desire for precise spatial registration.

User interaction patterns (Table~\ref{tab:editing-operations}) demonstrated effective spatial understanding and generation quality. Spatial understanding editing required predominantly rotation adjustments (M = 1.14) for scaffold alignment, with minimal semantic label corrections (M = 0.07). Generated content editing focused on position adjustments (M = 10.29 moves). Low regeneration frequency (M = 0.78) indicated satisfaction with initial style transfer quality.

The System Usability Scale score of 78.97 (SD = 13.83) places Roomify in the ``good'' to ``excellent'' range. Participants valued creative ownership despite time investment: \textit{``The time cost isn't problematic because this is my living space---I'll use it long-term. Plus, creating something myself gives me a sense of achievement''} (P18).

\subsubsection{Comparative Experience Evaluation}

\begin{figure*}[t]
\centering
\includegraphics[width=0.75\textwidth]{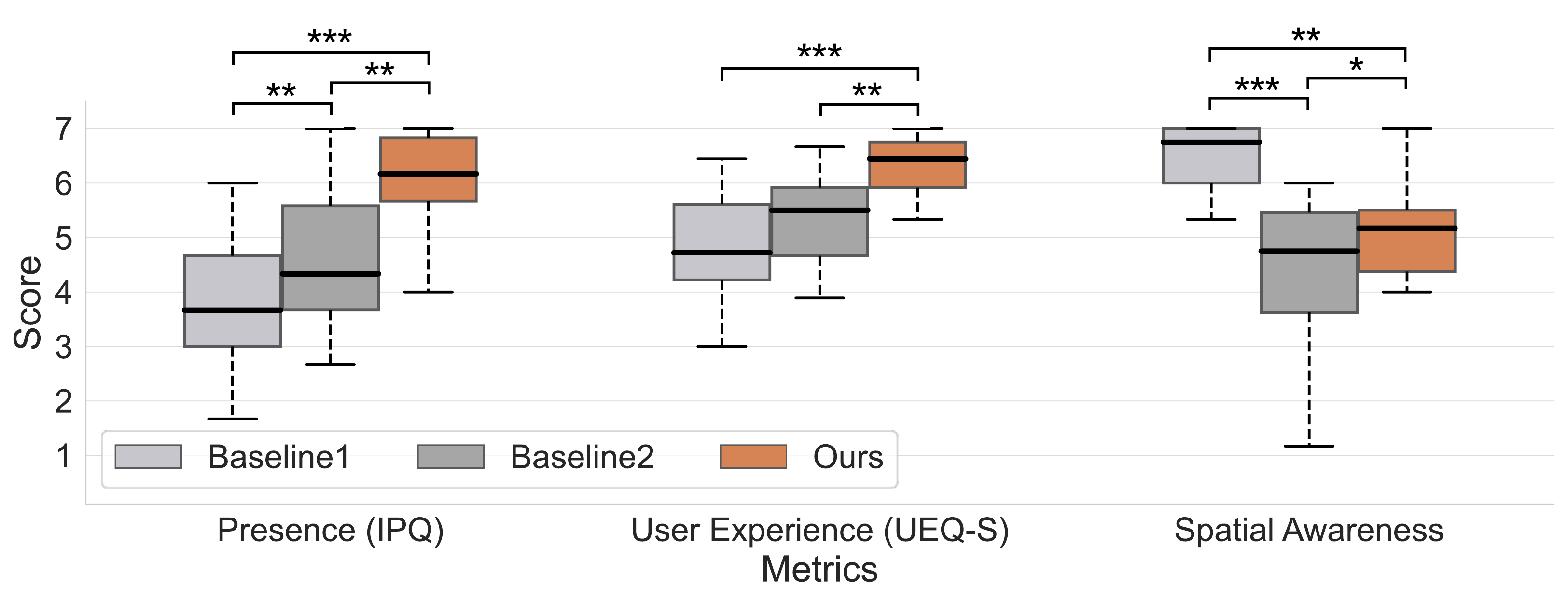}
\caption{Comparative evaluation across presence, user experience, and spatial awareness. Roomify achieves superior presence and experience quality while maintaining intermediate spatial awareness. Significance: * p < 0.05, ** p < 0.01, *** p < 0.001.}
\label{fig:metrics}
\end{figure*}

Analysis across conditions revealed significant differences in presence, experience quality, and spatial awareness (Fig.~\ref{fig:metrics}).

\begin{figure*}[t]
\centering
\includegraphics[width=0.75\textwidth]{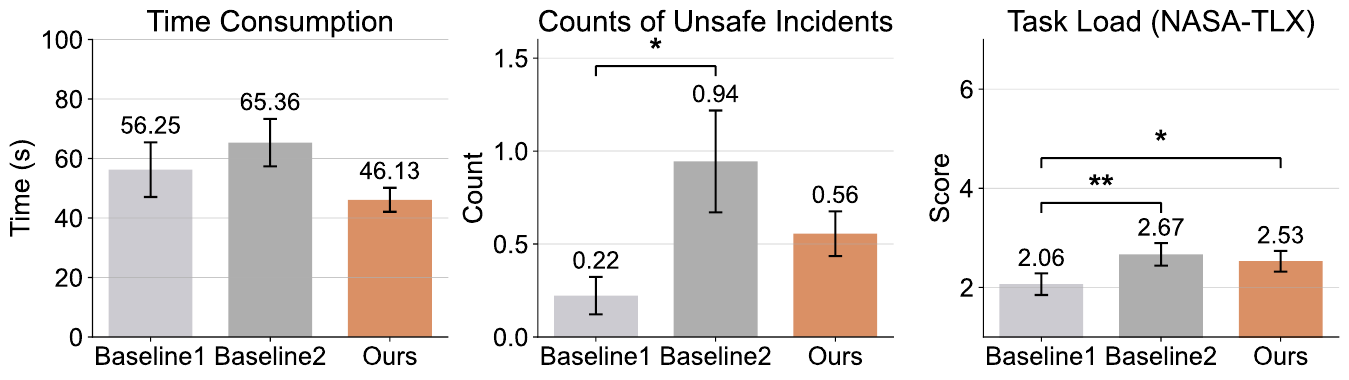}
\caption{Treasure hunt performance: (a) completion times, (b) spatial incident frequencies, (c) cognitive workload. Error bars show standard error. Significance: * p < 0.05, ** p < 0.01.}
\label{fig:performance}
\end{figure*}

\textbf{Presence and Immersion.} ANOVA revealed significant differences (F(1.91, 32.51) = 26.36, p < 0.001, $\eta_g^2$ = 0.38). Roomify achieved highest presence (M = 5.94, SD = 0.89), significantly outperforming Passthrough (M = 3.65, SD = 1.33, p < 0.001) and Fully Virtual (M = 4.72, SD = 1.40, p < 0.01)---representing 63\% and 26\% improvements respectively. Participants attributed this to thematic coherence: \textit{``The environment aligns perfectly with my movie's theme...the immersion from the surroundings and what I'm doing are thematically connected''} (P18).

\textbf{Experience Quality.} User experience showed significant variation (F(1.65, 28.02) = 21.22, p < 0.001, $\eta_g^2$ = 0.32), with Roomify scoring highest (M = 6.27, SD = 0.68) compared to Passthrough (M = 4.92, p < 0.001) and Fully Virtual (M = 5.27, p < 0.01). Participants emphasized integrated virtual-physical relationships: \textit{``Everything virtual around me is touchable...it feels magical. I truly feel I’m sitting in a spaceship seat''} (P12).

\textbf{Spatial Awareness.} Analysis revealed expected trade-offs (F(1.68, 28.60) = 16.85, p < 0.001, $\eta_g^2$ = 0.35). Passthrough provided maximum awareness (M = 6.20, SD = 1.17), exceeding both Roomify (M = 5.10, p < 0.01) and Fully Virtual (M = 4.41, p < 0.001). Importantly, Roomify maintained significantly better spatial awareness than Fully Virtual (p < 0.05), demonstrating that spatially-grounded transformation retains spatial understanding despite visual alteration: \textit{``Even though everything looks different, I still know where the sofa is---it's now an antique sofa''} (P15).

\begin{figure*}[t]
\centering
\includegraphics[width=\linewidth]{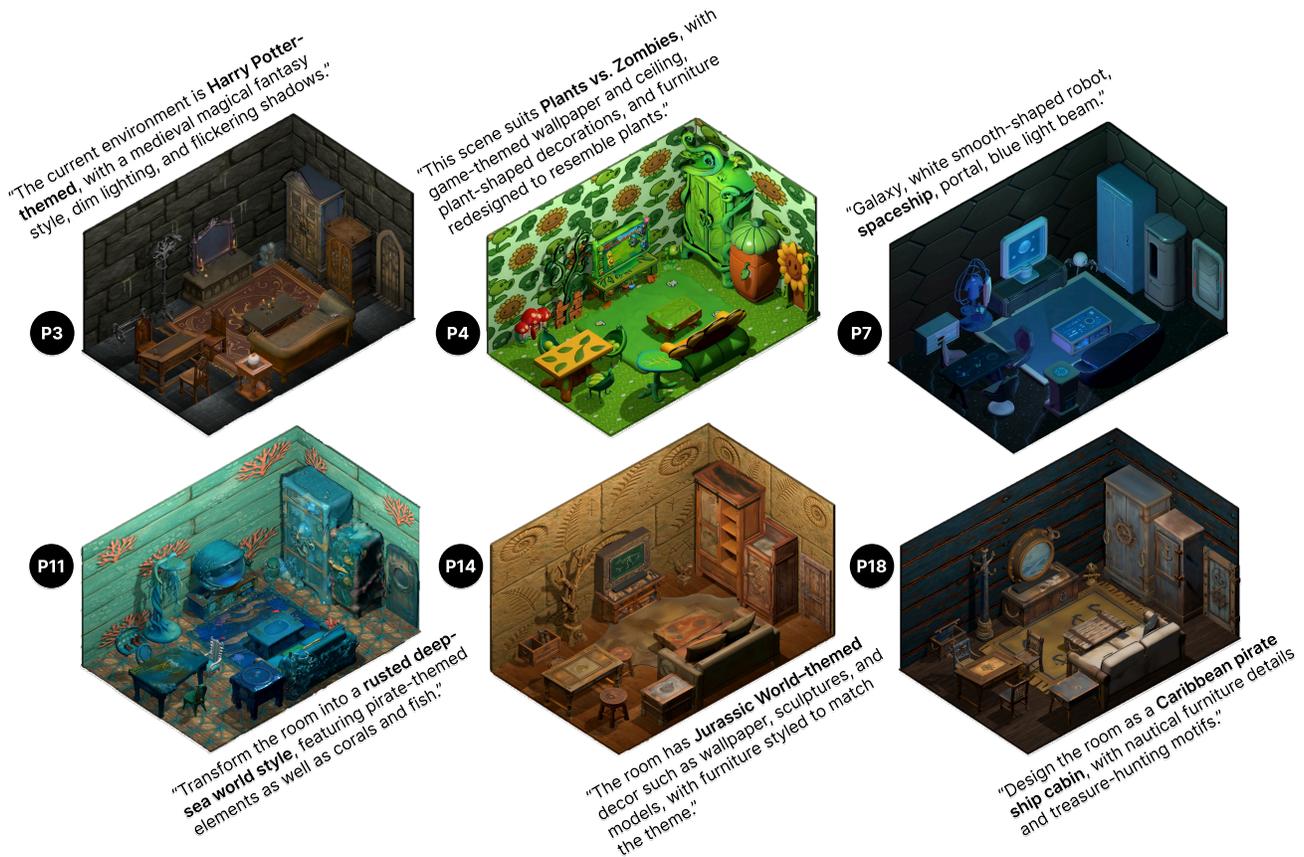}
\caption{Representative participant-created environments demonstrating successful style transfer across diverse themes while preserving spatial structure.}
\label{fig:creations}
\end{figure*}

Fourteen participants (78\%) selected Roomify as preferred. Figure~\ref{fig:creations} demonstrates successful style transfer across diverse themes. Participants reported enhanced connection through creative ownership: \textit{``Roomify creates a magical relationship with my space...a sense of ownership that makes me rethink the
possibilities of my static home environment''} (P18). The grounded transformation distinguished virtual experiences from pure fantasy: \textit{``It's not just fantasizing about my home becoming surreal...it's actually installed in real positions where I can see and touch it''} (P5).

\subsubsection{Navigation Performance}

Objective performance during the treasure hunt task revealed distinct navigation profiles (Fig.~\ref{fig:performance}).

\textbf{Task Efficiency.} While not statistically significant (p = 0.28), Roomify demonstrated fastest mean completion time (M = 46.13s) compared to Passthrough (M = 56.25s) and Fully Virtual (M = 65.36s). Participants attributed efficiency to reduced visual complexity: \textit{``Roomify hides irrelevant household items, making the environment cleaner''} (P1).

\textbf{Spatial Navigation.} Incident analysis revealed significant differences (F(1.97, 33.43) = 5.37, p < 0.01, $\eta_g^2$ = 0.12). Passthrough produced fewest incidents (M = 0.22), Fully Virtual generated most (M = 0.94, p < 0.05), with Roomify intermediate (M = 0.56). Roomify enabled predictable navigation: \textit{``I can see the whole room and anticipate furniture positions''} (P3).

\textbf{Cognitive Load.} NASA-TLX revealed significant differences (F(1.81, 30.83) = 7.23, p < 0.01). Passthrough imposed lowest load (M = 2.06), less than both Roomify (M = 2.53, p < 0.01) and Fully Virtual (M = 2.67, p < 0.05). Comparable workload between transformed conditions (p = 0.64) suggests navigating stylistically altered environments requires similar cognitive resources regardless of spatial preservation approach.

\textbf{Registration Accuracy Limitations.} Despite successful spatial structure preservation, registration precision emerged as a limitation. Occasionally, stylized geometry mismatches created uncertainty: \textit{``The coffee table was generated as oval-shaped, so I couldn't align it with the rectangular real table...making it hard to judge distances''} (P2). This highlights tension between artistic transformation freedom and precise spatial mapping for confident navigation.

\section{Study 2: Creative Prototyping Tool Evaluation}

\begin{table*}[t]
\centering
\caption{Study 2 participant demographics and professional design expertise.}
\label{tab:participants_study2}
\small
\begin{tabular}{cccllcc}
\toprule
\textbf{ID} & \textbf{Age} & \textbf{Gender} & \textbf{Education} & \textbf{Design Field} & \textbf{Exp. (yrs)} & \textbf{VR Familiarity} \\
\midrule
P1 & 23 & M & MSc in prog. & Architecture & 5 & Average \\
P2 & 23 & F & Ph.D. in prog. & Architecture & 5 & Average \\
P3 & 23 & F & Ph.D. in prog. & Architecture & 5 & Above Average \\
P4 & 25 & F & MSc in prog. & Architecture & 6 & Above Average \\
P5 & 22 & M & MSc in prog. & Architecture & 5 & Above Average \\
P6 & 27 & F & MSc in prog. & Visual/Science Art & 8 & Average \\
P7 & 19 & F & BSc in prog. & Product Design & 1 & Average \\
P8 & 23 & F & MSc & Film Storyboard & 2 & Average \\
\bottomrule
\end{tabular}
\end{table*}

\begin{figure*}[t]
\centering
\includegraphics[width=0.85\textwidth]{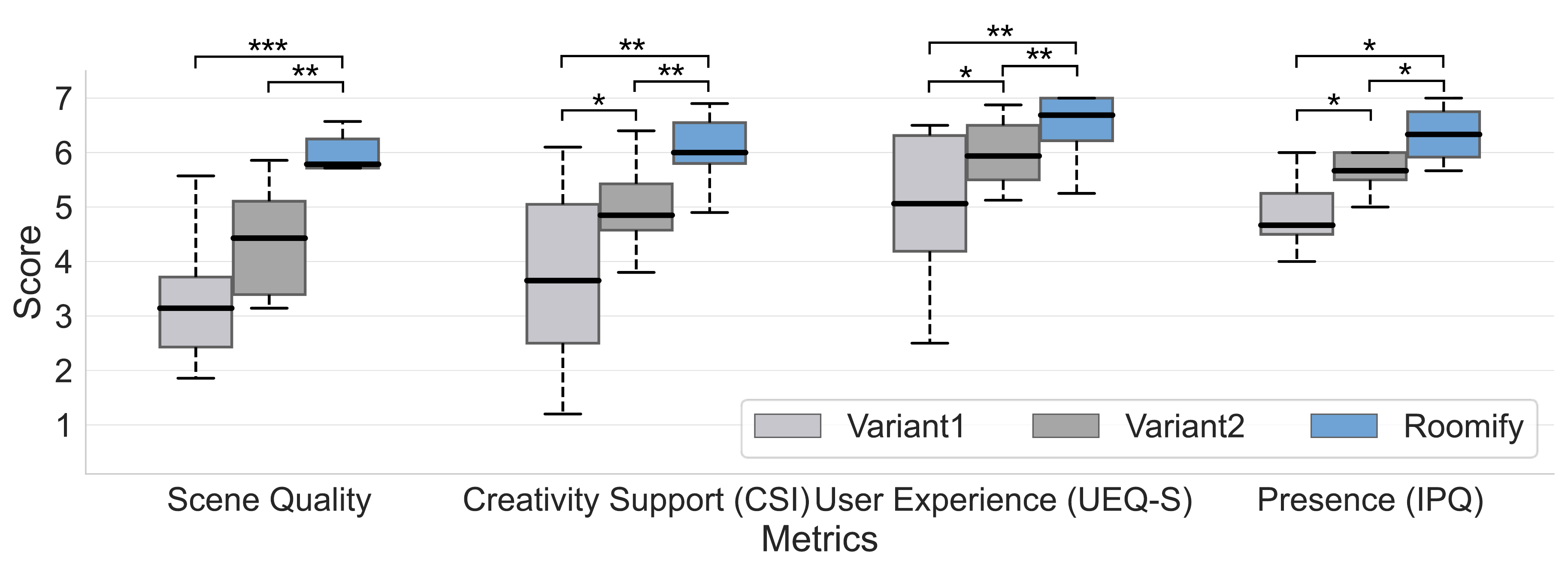}
\caption{Comparative performance across generation methods. Roomify demonstrates superior scene quality, creativity support, user experience, and presence. Error bars: standard error. Significance: * p < 0.05, ** p < 0.01, *** p < 0.001.}
\label{fig:study2_metrics}
\end{figure*}

Following our Formative Study and Study 1, participants frequently identified Roomify's potential as a creative prototyping tool, valuing its balance between generative capabilities, efficiency, and user control. They envisioned applications spanning interior design exploration, decorative theme previewing, and rapid concept iteration. To validate these observations, we conducted a focused evaluation with design professionals addressing two research questions:

\textbf{RQ3:} How do creative professionals evaluate the quality, expressiveness, and consistency of Roomify's generated prototypes?

\textbf{RQ4:} Can Roomify provide practical value as a prototyping tool for professional creative workflows compared to existing approaches?

\subsection{Experimental Design}

We employed a controlled ablation study comparing Roomify against two technical variants representing common VR scene generation approaches, isolating the contributions of spatially-grounded transformation.

\begin{itemize}
    \item \textbf{AI Re-Texturing (Variant 1)} applies style-consistent textures to scanned room geometry using Meshy's AI texture generator while preserving original object shapes. This represents surface-level stylization approaches \cite{yang2024dreamspace, song2023roomdreamer}.
    \item \textbf{Text-to-3D Generation (Variant 2)} generates objects directly from style prompts using Tripo v2.5 without reference to real object geometry. This represents standard text-to-3D pipelines \cite{zhang2024vrcopilot, hou2025echoladder}.
\end{itemize}

All variants utilized identical prompts from Roomify's mapping table, ensuring fair comparison of generation approaches.

\subsection{Participants}

We recruited 8 participants with professional design backgrounds (Table~\ref{tab:participants_study2}). All possessed experience with AI-assisted design tools. Architecture students constituted the majority (5/8), with remaining participants from visual arts, product design, and film production.

\subsection{Protocol and Tasks}

Each participant completed two creation tasks testing system versatility across spatial contexts:

\begin{itemize}
    \item \textbf{Real Environment Transformation:} Using the same physical living room from Study 1, participants created personalized themed environments through the complete Roomify pipeline.
    \item \textbf{Virtual Environment Transformation:} Participants transformed randomly assigned ScanNet \cite{dai2017scannet} environments (bedroom, classroom, office, or lounge) to test generalization across diverse spatial configurations.
\end{itemize}

For both tasks, participants exercised complete creative freedom in theme selection and prompt specification. During VR preview phases, participants could freely switch among the three generation variants using a dedicated controller button, with brief fade transitions between environments. We encouraged participants to take sufficient time experiencing and comparing conditions side-by-side before proceeding to questionnaires, ensuring ratings reflected informed within-subject comparisons.

Following task completion, participants completed standardized questionnaires and semi-structured interviews exploring authoring experience, comparative preferences with rationales, and potential integration within current design practices. Sessions required approximately 100 minutes; participants received \$25 compensation.

\subsection{Measurements}

We implemented multi-dimensional measurement: (1) System Usability Scale (SUS \cite{brooke2013sus}); (2) Scene Quality Scale (8 items) assessing style fidelity, detail quality, and spatial appropriateness; (3) Creativity Support Index \cite{cherry2014quantifying} (10 items); (4) User Experience Questionnaire-Short (UEQ-S \cite{hinderks2017design}); (5) Igroup Presence Questionnaire (IPQ \cite{schubert2001experience}). All measures used 7-point Likert scales. Statistical analysis utilized ART-ANOVA combined with thematic analysis of interview data.

\subsection{Results}

\begin{figure*}[t]
\centering
\includegraphics[width=\linewidth]{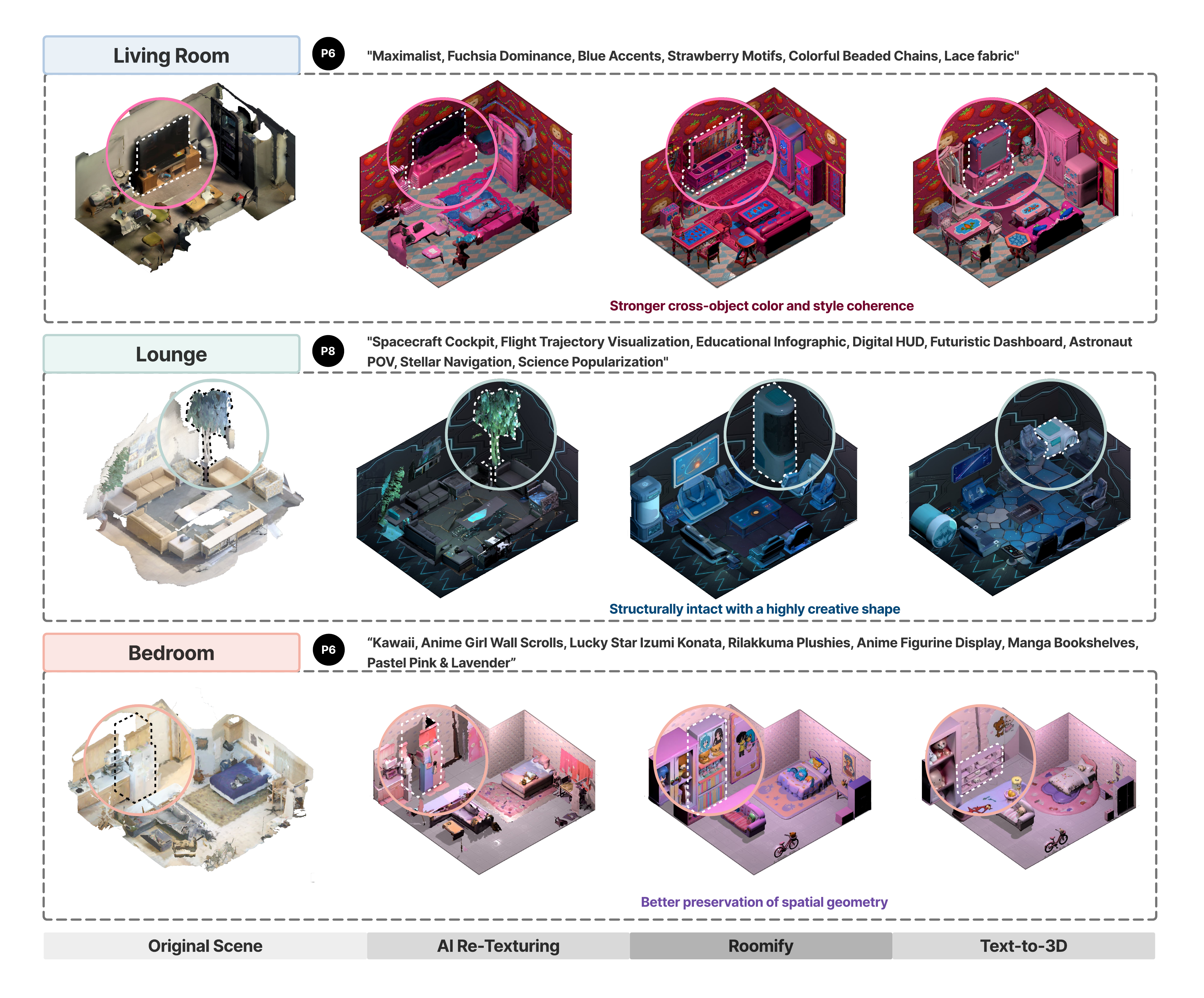}
\caption{Comparative generation results. Left: original environments. AI Re-Texturing shows limited stylization and mesh artifacts. Roomify demonstrates superior spatial consistency through reference-guided generation. Text-to-3D lacks spatial grounding, resulting in inconsistent placement. Examples show maximalist, spacecraft, and kawaii aesthetics.}
\label{fig:comparative_results}
\end{figure*}

\subsubsection{System Usability}

The System Usability Scale score of 84.38 (SD = 5.73) places Roomify in the ``excellent'' range. Participants appreciated the balanced automation: \textit{``It's quite intuitive and easy to get started''} (P7). Rapid generation emerged as a distinguishing strength: \textit{``The best part is its speed, and like all AI tools, it generates things beyond your expectations''} (P2).

\subsubsection{Comparative Generation Performance}

Systematic comparison revealed significant performance differences favoring Roomify (Fig.~\ref{fig:study2_metrics}).

\textbf{Scene Quality.} ANOVA revealed substantial differences (F(1.51, 10.59) = 23.07, p < 0.001, $\eta_g^2$ = 0.57). Roomify achieved highest ratings (M = 5.95, SD = 0.54), significantly exceeding AI Re-Texturing (M = 3.41, p < 0.001) and Text-to-3D (M = 4.50, p < 0.01). Figure~\ref{fig:comparative_results} illustrates quality differences. AI Re-Texturing showed limited stylization due to geometry constraints and mesh artifacts. Participants noted: \textit{``Roomify is best...the re-texturing looks broken and incomplete''} (P2). Although per-object quality for Roomify and Text-to-3D can appear comparable in some views, participants reported that Roomify provided stronger cross-object color and style coherence (e.g., in the maximalist living room and spacecraft lounge) and better preservation of spatial geometry (e.g., more accurate door, bed, and bookshelf proportions in the kawaii bedroom), which likely contributed to its higher overall scene-quality scores.

\textbf{Creativity Support.} Significant variation across methods (F(1.68, 11.73) = 19.83, p < 0.001, $\eta_g^2$ = 0.39), with Roomify scoring highest (M = 6.08, SD = 0.63) compared to AI Re-Texturing (M = 3.73, p < 0.01) and Text-to-3D (M = 5.09, p < 0.01). Participants valued creative inspiration: \textit{``It's completely different from what I imagined...this shows me how AI thinks, breaking barriers''} (P8). The generation of novel forms supported exploration: \textit{``I see forms beyond what design students typically encounter''} (P6).

\textbf{User Experience and Presence.} Both UEQ-S (F(1.08, 7.55) = 22.08, p < 0.01) and Presence (F(1.58, 11.05) = 11.43, p < 0.01) demonstrated Roomify's superiority. Immersive preview distinguished the system: \textit{``It feels like playing a game...there's a sense of being there''} (P5). Rapid iteration enhanced creative flow: \textit{``The quick visualization of real states is invaluable''} (P6).

\begin{figure*}[t]
\centering
\includegraphics[width=\textwidth]{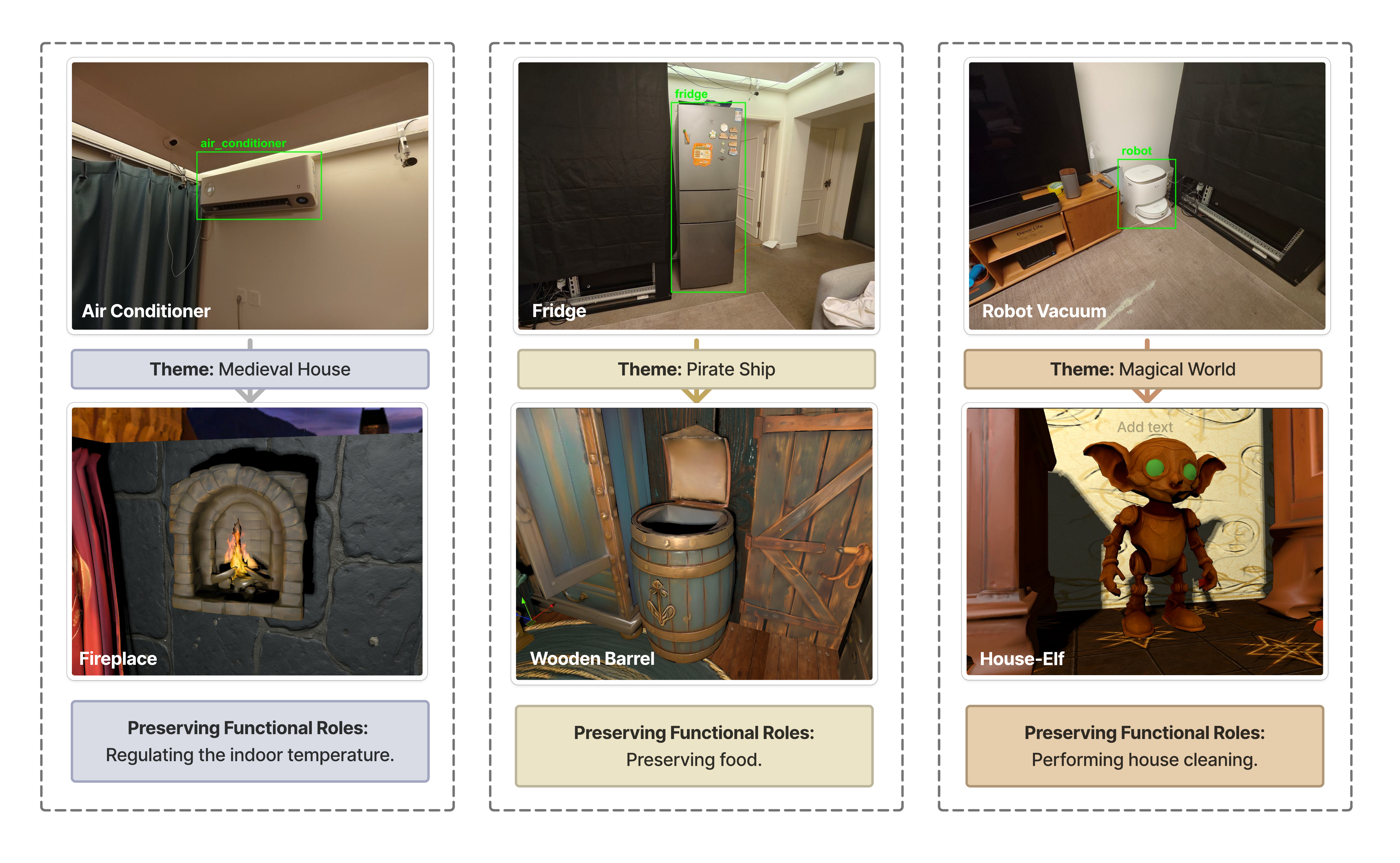}
\caption{functional semantics consistency examples. The system preserves object functional roles while achieving thematic integration: an air conditioner becomes a fireplace (temperature regulation) in a medieval house, a refrigerator transforms into a barrel (storage) on a pirate ship, a cleaning robot reimagines as a house-elf (cleaning) in the magical world.}
\label{fig:functional_consistency}
\end{figure*}

\subsubsection{Functional Consistency and Creative Innovation}

A distinctive strength of Roomify is that it combines style diversity with functional consistency and geometric preservation. Participants often described the results as ``playful reskins'' of the room where the overall structure and functions remained understandable despite dramatic visual changes. As shown in Figure~\ref{fig:functional_consistency} , an air conditioner can become a medieval fireplace that still occupies a similar wall area and is interpreted as part of the room’s temperature regulation, and a refrigerator can transform into a pirate barrel that preserves the idea of a storage container. Participants recognized this as a creative advantage: \textit{``The system doesn't just change how things look---it understands what they do and finds creative ways to preserve that within the new theme''} (P4). This semantic awareness enabled design explorations maintaining recognizable object semantic roles while supporting radical aesthetic transformation and achieving thematic coherence.

In these cases, Roomify preserves functional semantics (e.g., seating, storage, support) and gross geometry (approximate volume and footprint), which is sufficient for users to navigate, avoid collisions, and treat objects as sit-able or storable surfaces. However, we explicitly do not claim that all fine-grained affordances are preserved: in the refrigerator$\rightarrow$barrel example, users may still understand the barrel as storage and avoid walking through it, but refrigeration, door mechanics, and other detailed action possibilities are not replicated. 


\subsubsection{Professional Workflow Integration}

All eight participants selected Roomify as their preferred method, identifying specific application scenarios:

\textbf{Early-Stage Conceptual Development.} Participants positioned Roomify within initial design phases: \textit{``It's most valuable in initial concept stages and style determination''} (P5), functioning as a \textit{``brainstorming tool for rapid conceptual generation''} (P6).

\textbf{Client Communication.} Architecture professionals identified value for client interactions: \textit{``It's perfect for architect-client dialogue...clients need intuitive visualization to confirm desired styles''} (P4). Immersive preview enables experiential communication where clients inhabit proposed spaces rather than interpret abstract representations.

\textbf{Media Production.} Entertainment participants recognized storyboarding applications: \textit{``We can use it for layout and storyboard testing...placing cameras inside to test character-scene relationships''} (P8).

\textbf{Emerging Applications.} Participants envisioned theatrical stage design (P7), educational spatial imagination development (P2), and virtual exhibition creation (P1).

\begin{figure*}[t]
    \centering
    \includegraphics[width=\textwidth]{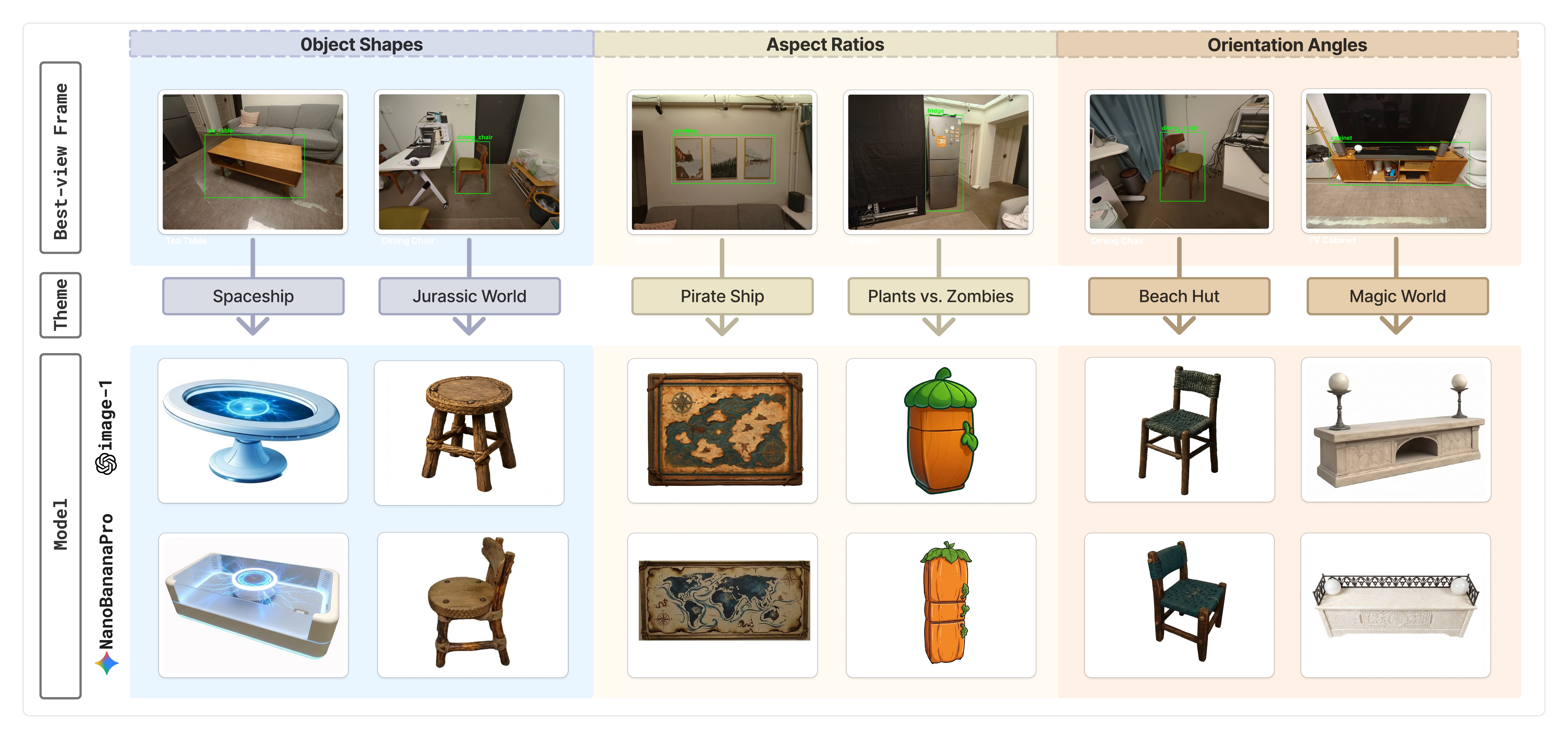}
    \caption{Geometric hallucination failure cases and model comparison. The second-to-last row shows GPT Image-1 outputs with three hallucination types: shape distortion (rectangular table becomes round), aspect ratio changes (vertical frame becomes horizontal), and orientation errors (chair faces different direction). The bottom row shows improved outputs from Google's Gemini 3 Pro Image (NanoBanana Pro), demonstrating substantially reduced geometric deviations across all three categories. 
    }
    \label{fig:failure_cases}
\end{figure*}

\subsubsection{Limitations}

Design professionals demonstrated realistic understanding of constraints. Generated models require refinement for production: \textit{``The model accuracy can't be directly used in design---requires manual detail adjustment''} (P2). Style modification is limited to regeneration rather than parametric adjustment: \textit{``Users can't adjust object shapes or colors directly, only through regeneration''} (P5). Future iterations should incorporate fine-grained control mechanisms enabling direct manipulation of object properties such as materials, textures, colors, or geometric details.

Participants also recognized Roomify's niche as rapid conceptual exploration rather than production asset generation. The system's strength lies in early-stage exploration and client communication, complementing rather than replacing traditional methods.

\section{Discussion}

Through two complementary studies, we establish key insights about virtual environment transformation and validate our approach's unique advantages.

\subsection{Reconciling Immersion and Spatial Awareness}

While video passthrough solutions enable detailed object manipulation and maximize spatial awareness \cite{hartmann2019realitycheck, wang2022realitylens, xiong2024petpresence}, they fundamentally break immersion by overlaying camera feeds unrelated to the user's virtual experience. Our approach transforms spatial cues into style-consistent virtual elements rather than breaking immersion through non-diegetic overlays. Study 1 reveals that Roomify achieves superior presence (63\% improvement over passthrough, 26\% over fully virtual) while maintaining significantly better spatial awareness than fully virtual environments with proximity-based boundary display. This demonstrates that spatial awareness can be achieved while enhancing immersion by making the physical environment an integral, enriching part of the virtual experience.

\subsection{Generation Approach and Failure Cases}

Rather than attempting direct stylization of complex 3D scenes, Roomify employs a strategic ``dimension reduction then elevation'' approach. The system first selects representative best-view frames for each object, leverages GPT Image's spatial understanding and style application capabilities to generate 2D stylized images, then uses these intermediate results to guide 3D object generation. This approach combines stylistic ``essence'' with spatial ``structure,'' and Study 2 confirms that compared with text-to-3D methods, Roomify achieves superior spatial geometry consistency and cross-object style coherence. This 2D-to-3D strategy has been adopted by related work including ImaginateAR \cite{lee2025imaginatear} and WorldGen \cite{wang2025worldgen}, as foundation models demonstrate stronger understanding and generation capabilities for images than for 3D models due to more abundant training data.

However, due to current limitations of multimodal large language models, reference-based image generation still exhibits geometric hallucinations. Our user studies indicate that hallucination remains prevalent in current generative pipelines, particularly in preserving: (1) \textbf{object shapes}---models may transform rectangular tables into circular ones; (2) \textbf{aspect ratios}---wardrobes may have altered proportions; and (3) \textbf{orientation angles}---chairs may face different directions than their physical counterparts. Figure~\ref{fig:failure_cases} illustrates these typical failure cases.

For VR users requiring physical movement, spatial and geometric consistency maximizes navigation confidence and safety. To mitigate the impact of misaligned geometry, our Cross-Reality Authoring Tool integrates an MR correction mechanism enabling users to rapidly adjust generated object transforms (Fig.~\ref{fig:authoring_overview}), with collision-risk objects (tables, chairs, large furniture) red-flagged for position confirmation. In Study 1, participants performed an average of 16.93 adjustments on generated content (10.29 moves, 4.07 rotations, 2.57 scales), reflecting the operational burden caused by limited geometric consistency. However, the adjustment functionality cannot resolve all hallucination-induced issues---shape deformations cannot be corrected through simple transform adjustments. When a rectangular table is generated as circular, users may misjudge distances during movement, potentially causing collisions.

Encouragingly, as multimodal foundation models rapidly evolve, reference-based control capabilities have significantly improved. In recent pilot experiments, we found that newer generation models---Google's Gemini 3 Pro Image\footnote{https://deepmind.google/models/gemini-image/pro/} (NanoBanana Pro, released November 2025)---demonstrated substantial improvements in generation quality compared to OpenAI's GPT Image-1 (released April 2025) used in our studies, exhibiting far fewer geometric deviations and enhanced spatial structure stability. As shown in Figure~\ref{fig:failure_cases}, Gemini 3 Pro Image significantly reduces alignment errors in object shapes, aspect ratios, and orientation angles. As multimodal models continue advancing, these hallucination issues will likely further diminish, reducing user operational burden and safety risks while improving Roomify's usability. Future work could also explore improving mesh-based or 3DGS-based 3D-to-3D generation methods \cite{hollein2022stylemesh, yang2024dreamspace, liu2024stylegaussian} to reduce geometric alignment hallucinations introduced during the dimension reduction process.

\subsection{Balancing Geometric Fidelity and Creative Freedom}

Our evaluation reveals a nuanced relationship between user intent, spatial constraints, and transformation expectations. The unpredictability of AI generation represents both an opportunity and a challenge, with different user contexts requiring different balance points between geometric fidelity and creative freedom.

For VR users requiring physical movement, our safety benefits primarily arise from geometric preservation of obstacles and walkable corridors. By keeping the positions, sizes, and major contact surfaces of furniture consistent between the physical and virtual scenes, Roomify helps users avoid tripping and collisions and supports gross body-support actions such as sitting or leaning. In this case, complete spatial and geometric consistency maximizes navigation confidence and safety.

However, for creative users or static viewing contexts, geometric flexibility enables transformations that transcend surface appearance to achieve deeper thematic integration. The semantic transformation examples in Study 2 (Figure~\ref{fig:functional_consistency}) demonstrate how semantic understanding can guide appropriate geometric modifications that enhance rather than compromise design intent.

We propose that future systems should adaptively balance stylistic freedom and geometric anchoring based on usage context. For movement-intensive applications, safety-critical objects should maintain strict geometric and structural consistency. For static viewing or artistic design purposes, geometric constraints can be relaxed to enable greater stylistic freedom and creative expression.

\section{Limitations and Future Work}

While our evaluation demonstrates Roomify's effectiveness in balancing immersion with spatial awareness, several limitations inform directions for future development.

\textbf{Static Environment Assumption.} The current system assumes static indoor furniture and does not handle dynamic interactions. The system provides spatial cues enabling users to approximately know where they can sit or walk through geometric and semantic alignment, but true force feedback and tangible interactions \cite{fan2025tangiar} like opening doors or appliance controls are not fully modeled. Additionally, single-scan spatial understanding cannot accommodate continuously moving entities such as people or pets. Future development should integrate real-time spatial tracking and virtual replica methods like VirtualNexus and GaussianNexus \cite{huang2024virtualnexus, huang2024surfshare, huang2025gaussian} to handle dynamic scene elements.

\textbf{Evaluation Scope.} Both studies recruited relatively young participants who may exhibit stronger preferences for novel experiences compared to broader populations. Furthermore, our evaluation focused exclusively on individual experiences, leaving multi-user applications unexplored. The system's effectiveness for collaborative scenarios \cite{numan2024spaceblender}, multi-user communication \cite{hu2024thing2reality}, and social VR contexts \cite{sykownik2023vr, wang2025vr} remains unvalidated.

\textbf{Baseline Design in Study 1.} Our Study 1 baselines were designed to balance immersion and safety, which may introduce biases. The Fully Virtual condition showed only skybox without virtual furniture objects, and the boundary system remained active throughout. These design choices, while necessary for safety and fair comparison of spatial grounding mechanisms, may have conservatively estimated immersion in the Fully Virtual condition. Future studies should explore more balanced baselines that better isolate the effects of spatially-grounded transformation.

\section{Conclusion}
Virtual reality systems have long faced a fundamental design tension between immersive experiences and spatial awareness. This work presents Roomify, a spatially-grounded transformation system that unifies real-world context, AI-driven creativity, and user-driven personalization into one seamless experience. Our evaluation with 18 VR users and 8 design professionals demonstrates that spatially-grounded transformation enhances presence (63\% improvement over passthrough, 26\% over fully virtual) while maintaining reasonable spatial awareness. Design professionals validated the system's value for creative workflows including interior design exploration, client visualization, and media storyboarding, with high ratings for scene quality (5.95/7) and creativity support (6.08/7). 

By making physical environments integral to virtual experiences rather than obstacles to circumvent, Roomify enables users to remain ``in the story'' while maintaining spatial intuition. This opens new possibilities for domestic VR applications---from themed entertainment to professional design prototyping---where immersive computing enhances rather than replaces our relationship with physical spaces.

\begin{acks}
This work is supported by the Natural Science Foundation of China (NSFC) under Grant No. 62132010.
\end{acks}

\bibliographystyle{ACM-Reference-Format}
\bibliography{References}

\appendix

\section{Algorithms}
\subsection{Per-Object Best-View Frame Selection Algorithm}

This algorithm is designed to determine, for each in-scene object, the single best-view frame that serves both as the reference for stylized content generation and as the orientation anchor during registration. It evaluates frames according to three criteria: visibility, centering, and occlusion. Visibility ensures that sufficient object surface is observed, centering emphasizes perceptual salience by keeping the object near the image center, and occlusion filtering eliminates views where objects are significantly blocked. The lexicographic priority—visibility first, followed by centering, then visible area—ensures the selected frame maximizes clarity, perceptual balance, and contextual validity.

The procedure begins by transforming semantic scaffolds into SLAM coordinates using Sim(3) alignment. Each object’s corners are then projected into all camera frames, with occlusion tests applied to discard obstructed points. For each frame, a composite score is computed from visible point count, distance to the image center, and visible area. The best-view frame is selected via lexicographic maximization. Finally, the chosen frame is saved with a green bounding box and label rendered for visualization, and its associated camera pose is written into the global scene JSON. This enables both style-aware content generation by the agent and orientation placement during registration.

\begin{algorithm}[htbp]
\small 
\caption{Per-Object Best-View Frame Selection}
\KwIn{
$T^{c}_{f,\mathrm{SLAM}} \in \mathbb{R}^{4\times 4}$: per-frame camera extrinsics in SLAM coordinates; \\
\quad
$K \in \mathbb{R}^{4\times4}$: camera intrinsics; \\
\quad
Scene JSON (world coordinates): Semantic decomposition output JSON; \\
\quad
$T^{\mathrm{sim3}}_{\mathrm{world}\leftarrow \mathrm{SLAM}}=\{R_s,t_s,s\}$: Sim(3) alignment transform 
($R_s\in SO(3)$, $t_s\in \mathbb{R}^3$, $s>0$).
}
\KwOut{
For each object $o_i$: best frame index $f_i^\star$; rendered frame $I_{f_i^\star}$ with 2D green bounding box and label; 
original JSON updated with 
$\texttt{best\_frame\_pose}=T^{c}_{f_i^\star,\mathrm{SLAM}} \in \mathbb{R}^{4\times4}$.
}

\BlankLine
\textit{/* Geometry mapping: World $\to$ SLAM */}

\ForEach{$o_i \in \mathcal{O}$}{
  $c_i^s \gets \tfrac{1}{s} R_s^\top (c_i^w - t_s)$; \\
  $d_i^s \gets \tfrac{1}{s} d_i^w$; \\
  $\psi_i^s \gets \mathrm{atan2}((R_s^\top e_x^w)_y,(R_s^\top e_x^w)_x)$, 
  where $e_x^w=R_z(\psi_i^w)[1,0,0]^\top$; \\
  Construct 8 corners $v_{i,k}^s$ from $(c_i^s,d_i^s,\psi_i^s)$. \\
}

\BlankLine
\textit{/* Per-frame projection and occlusion */}

\ForEach{frame $f \in \mathcal{F}$}{
  \ForEach{object $o_i$}{
    Project corners: $X^c_{i,k}(f) = R^{(f)}_{cw,s} v_{i,k}^s + t^{(f)}_{cw,s}$; \\
    Pixel coords: $\pi([x,y,z]) = [f_x x/z + c_x, f_y y/z + c_y]^\top$, $z>0$; \\
    Keep valid points: $\mathcal{C}_i(f) = \{k \mid z>0, 0<u<W, 0<v<H\}$; \\
    Apply occlusion test against all occluders $o_j$: \\
    \Indp
    If $\pi(X^c_{i,k}(f)) \in \mathrm{Hull}_j(f)$ and 
    $z^{\text{near}}_j(f)+\delta_z^s < X^c_{i,k}(f)_z$, then discard point. \\
    \Indm
    Remaining set: $\mathcal{C}_i^+(f)$. \\
  }
}

\BlankLine
\textit{/* Frame-level scoring */}

\ForEach{$o_i$}{
  \ForEach{$f \in \mathcal{F}$}{
    If $\mathcal{C}_i^+(f)=\varnothing$, $s_i(f) \gets -\infty$; \\
    Else compute: \\
    \Indp
    $\text{vis\_cnt}_i(f)=|\mathcal{C}_i^+(f)|$; \\
    $\text{center\_uv}_i(f)=\tfrac{1}{|\mathcal{C}_i^+(f)|}\sum_{k\in\mathcal{C}_i^+(f)}\pi(X^c_{i,k}(f))$; \\
    $\text{center\_dist}_i(f)=\|\text{center\_uv}_i(f)-[W/2,H/2]^\top\|_2$; \\
    $\text{vis\_area}_i(f)=\mathcal{A}(\mathrm{Conv}\{\pi(X^c_{i,k}(f))\}_{k\in\mathcal{C}_i^+(f)})$; \\
    $s_i(f)=(\text{vis\_cnt}_i(f), -\text{center\_dist}_i(f), \text{vis\_area}_i(f))$. \\
    \Indm
  }
  Select best: $f_i^\star = \arg\max_{f\in\mathcal{F}} s_i(f)$ (lex order). \\
}
\end{algorithm}

\subsection{The Generated Content Registration Algorithm}

This algorithm is designed to integrate the Generated Content into a cohesive virtual environment. The registration process begins with unified scaling normalization, where each generated object undergoes isotropic scaling to match the longest edge of its corresponding semantic scaffold. This preserves proportional relationships while ensuring objects fit within their designated spatial boundaries. Orientation optimization follows, generating candidate rotations based on best-view camera poses and selecting optimal orientations through Intersection-over-Union (IoU) maximization between object and scaffold geometries.

For planar objects such as carpets or wall-mounted elements, the algorithm extends beyond simple yaw rotation to test axis flips, addressing cases where rotational adjustments alone cannot recover correct spatial orientation. This approach leverages the spatial constraints established during generation: since stylized objects maintain orientation consistency with their best-view frames, and best-view frames align with best-view poses, the camera pose provides natural orientation constraints that preserve user spatial intuition.

Fine-grained scaling refinement ensures object dimensions along each axis remain within $1.3\times$ scaffold extents, an empirically chosen threshold that balances alignment fidelity with stylistic flexibility, preventing spatial conflicts while allowing reasonable geometric variation. Ground plane alignment positions each object's bottom surface to match its scaffold's base, ensuring stable spatial grounding and consistent floor plane relationships throughout the environment.

\begin{algorithm}[htbp]
\small 
\caption{Generated Content Registration}
\KwIn{Scene JSON with objects $\{o_i\}_{i=1}^N$,\\
\hspace{1.8em}where each $o_i$ includes its scaffold $M_i$ and best-view camera yaw $\theta^\star_{y,i}$.}
\KwOut{Placed objects with optimized orientation and scale}

\ForEach{$o_i$}{
  \uIf{$o_i \in$ general $\;\cup\;$ doors/windows}{
    \tcp{(A) Unified longest-edge alignment (isotropic)}
    Compute model OBB longest edge $L_{\text{model},i}$ and scaffold's longest edge $L_{M,i}$;\\
    Set isotropic scale $s_i \leftarrow L_{M,i} / L_{\text{model},i}$ and apply $s_i$ to $o_i$;\\[0.25em]

    \uIf{$o_i \in$ general}{
    \tcp{if shortest edge $< 0.15 \times$ longest edge (thin object)} 
      \uIf{$o_i \in$ flat}{        
        \tcp{(B1) Flat-specific axis-flip search}
        Test axis flips $\{R_x(90^\circ),\,R_y(90^\circ),\,R_z(90^\circ)\}$;\\
        Keep flip $R_{\text{flip}}$ maximizing $\mathrm{IoU}$ (defined below);\\
        Apply $R_{\text{flip}}(\theta^\star)$;\\
      }
      \Else{
        \tcp{(B2) General yaw-only search around best-view yaw}
        Generate yaw set $\Theta_i = \{\theta : \theta \in [\theta^\star_{y,i}-45^\circ,\;\theta^\star_{y,i}+45^\circ],\;\Delta\theta=5^\circ\}$;\
        $\theta^\star \leftarrow \arg\max_{\theta \in \Theta_i}\mathrm{IoU}\!\left(B_i(R_y(\theta)),\,M_i\right)$;\\
        Apply $R_y(\theta^\star)$;\\
      }
    }
    \ElseIf{$o_i \in$ doors/windows}{
      \tcp{(B3) Door/window logic}
      Align in-plane orientation to host-wall direction;\\
      Fix thickness to constant $t_0$;\\
    }

    \tcp{(C) Refine scaling with scaffold guard}
    Refine per-axis scale so that extents do not exceed $1.3\times$ those of $M_i$;\\
    Align the model’s bottom face with the bottom face of $M_i$;\\
  }
  \ElseIf{$o_i =$ wall}{
    \tcp{Walls: placement + texture}
    Place per semantic segmentation; shift outward by $5$\,cm to avoid occluding doors/windows;\\
    Apply wall PBR texture;
  }
  \ElseIf{$o_i =$ floor}{
    \tcp{Floor: span and texture}
    Span the $x$--$z$ extent bounded by walls; apply floor PBR texture;
  }
  \ElseIf{$o_i =$ skybox}{
    \tcp{Skybox: load dynamic panoramic skybox}
    Load generated skybox; Loop the dynamic skybox video for seamless playback;\\
  }
}
\end{algorithm}

\section{Prompts Used in the Pipeline}

This section presents several core prompts used in the Roomify Pipeline, provided for reproducibility and methodological transparency, and illustrating how task-specific instructions are designed and applied across different stages of the system.

\subsection{Style Extraction Prompt}
\label{app:style-extraction}

\begin{lstlisting}[language=Python, caption=Style Extraction, label=lst:style-extractor-img]
style_extraction_prompt = ChatPromptTemplate.from_messages(
    [
        (
            "system",
            "You are a professional style-extraction assistant focused on virtual scene generation. "
            "An image describing the user's intended task has been uploaded. Your job is to extract 4-8 precise "
            "English style keywords from either the user's text or the image description. "
            "For example, if the user says 'I want to turn the room into a Cyberpunk style', your extracted keywords "
            "must include 'Cyberpunk'. The keywords should describe materials, colors, eras, or architectural features. "
            "Avoid vague words (e.g., 'beautiful'). If the input is unclear, return the default style 'Modern Minimalist'. "
            "For non-English inputs, first map them to English. Separate the phrases with commas."
            "Examples:"
            "User input: I want to make the room in Plants vs. Zombies style; Expected output: Plants vs. Zombies, Cartoonish, Whimsical, Bright Green, Wooden Fence, Vibrant Colors, Playful Garden, Pop Art"
            "User input: Caribbean Pirate style; Expected output: Pirates of the Caribbean, Nautical, Rustic Wood, Weathered Canvas, Aged Bronze, Dark Ocean Blue, Caribbean Colonial, Medieval Ship"
            "User input: Cyberpunk style; Expected output: Cyberpunk, Neon Lights, Chrome Metal, Electric Blue, Hot Pink, Futuristic, High-tech, Dystopian Urban"
            "User input: I want Dark Gothic style, but the curtains should be white; Expected output: Dark Gothic, Black Stone, Ironwork, Candlelight, Stained Glass, White Curtains, Medieval Architecture, Dramatic Shadows"
        ),
        ("human", "{user_sentence}"),
    ]
)
\end{lstlisting}

\subsection{Mapping Table Inference Prompt}
\label{app:Mapping Table}
\begin{lstlisting}[language=Python, caption=Mapping Table Inference, label=lst:Mapping-Table-Inference]
mapping_example_prompt = PromptTemplate(
    input_variables=["style", "objects", "output"],
    template=(
        "[Example]\n"
        "Target style: {style}\n"
        "Detected real-world objects (object_id:label): {objects}\n"
        "Expected output (JSON object, fields: objects, skybox, wall_texture, floor_texture):\n{output}\n"
        "--"
    ),
)

mapping_prompt = FewShotPromptTemplate(
    examples=vr_styliser_few_shot_examples,
    example_prompt=vr_styliser_example_prompt,
    prefix=(
        "You are a professional VR scene design assistant. Your task is to replace real-world objects "
        "with virtual replicas that match the target style.\n"
        "## Reasoning Sequence\n"
        "First, based on the semantic label information of objects provided in the JSON file, infer each object's "
        "object_function. Then, find a replica with the same function according to the style keywords. Finally, "
        "using both the style and the spatial information contained in the JSON file (coordinates, position, size, etc.), "
        "generate an appearance_prompt that describes the replica's overall appearance (mainly color and material), "
        "and infer its safety collision_risk.\n"
        "## Output Format\n"
        "Return a **JSON object**, example structure:\n"
        "{\n"
        "  \"objects\": [ [ ...7 columns... ], ... ],\n"
        "  \"skybox\": {\"prompt\": \"...\", \"negative_text\": \"...\"},\n"
        "  \"wall_texture\": {\"prompt\": \"...\"},\n"
        "  \"floor_texture\": {\"prompt\": \"...\"}\n"
        "}\n"
        "Field Description:\n"
        "- \"objects\": 2D array, each row in order: object_id, label, object_function, replica, replica_function, appearance_prompt, collision_risk\n"
        "- \"skybox\": skybox prompt object\n"
        "- \"wall_texture\": seamless wall texture prompt object\n"
        "- \"floor_texture\": seamless floor texture prompt object\n"
        "## Requirements\n"
        "- Ensure materials, colors, and textures are consistent with the target style\n"
        "- For wall/floor textures, descriptions must be seamlessly tileable and consistent with in-scene objects, "
        "without being overly eye-catching\n"
        "- The appearance_prompt must consider object size, camera position, object center, and rotation angles; "
        "size must match the original object, but no numeric values should appear\n"
        "- The appearance_prompt should be detailed (100-200 words), controlling shape, material, color, and texture\n"
        "- For ambiguous labels, reasonably infer their function\n"
        "- collision_risk should only be true/false. Mark true if the object is likely to be physically contacted "
        "by the user in the room (e.g., large furniture, tables, sofas, beds, chairs). Mark false for curtains, windows, "
        "doors, or other wall-adjacent/flat/soft objects\n"
        "**Only return JSON, without any explanation or extra text.**"
    ),
    suffix=(
        "## Task Start\n"
        "User's expected style keywords: {style}\n"
        "Detected real-world objects (object_id:label): {objects}\n"
        "Scene JSON information containing bbox positions, dimensions, etc.: {scene_json}\n"
        "Please generate the complete JSON:"
    ),
    input_variables=["style", "objects", "scene_json"],
)
\end{lstlisting}

\subsection{Stylized Image Generation Prompt}
\label{app:Image Generation}
\begin{lstlisting}[language=Python, caption=Stylized Image Generation, label=lst:Stylized-Image-Generation]
def build_image_prompt(obj: Dict, scene_data: Dict) -> str:
    """
    Constructs the stylized image generation prompt.

    - obj['label']            -> Original object name
    - obj['object_function']  -> Original object function
    - obj['replica']          -> Replica name
    - obj['replica_function'] -> Replica function
    - scene_data['style']     -> Global style for the scene
    """
    label = obj['label'].lower()
    is_surface = label in ['wall', 'floor', 'ceiling']

    # Extract style_prompt from the object if available, otherwise fall back to global style
    style_prompt = scene_data.get('style', 'Modern Minimalist')
    size_req = ""
    if obj.get("size"):
        size_req = (
            f"The object's size is {obj['size']} ([x,y,z], z-up), and its yaw rotation angle in the scene "
            f"is {obj['rotation']} rad. Please ensure that the generated stylized substitute strictly preserves "
            f"this scale and takes the rotation into account. "
        )

    return (
        "You are an image-stylization assistant. "
        f"Transform the object in the uploaded image, which serves as {obj['object_function']} "
        f"({obj['label']}), into a {obj['replica']} that conforms to the {style_prompt} style "
        f"and performs the {obj['replica_function']} function. "
        "Render the final image from a 45-degree perspective, under neutral lighting, and leave the background blank. "
        f"Details: {obj['prompt']}. "
        f"{size_req}"
        "**Additional requirements**: The output must be a PNG with a **transparent background**. "
        f"Focus exclusively on the target object ({obj['label']})-the region marked by the green bounding box-"
        "and disregard all other content in the image. The stylized output must replicate the exact angle, "
        f"dimensions, and proportions of the reference object, as well as the provided numerical data ([x,y,z], z-up: {obj['size']}). "
        "Slight variations in shape are acceptable. Provide the complete stylized result without any occlusion."
    )
\end{lstlisting}

\subsection{Dynamic Skybox Generation Prompt}
\label{app:Skybox Generation}
\begin{lstlisting}[language=Python, caption=Dynamic Skybox Generation, label=lst:Dynamic-Skybox-Generation]
image_to_video_prompt = ChatPromptTemplate.from_messages(
    [
        (
            "system",
            "You are a professional video generation assistant. Your task is to convert a static image into a dynamic video.\n"
            "## Core Requirements\n"
            "- The camera must remain completely still, with no panning, zooming, or movement\n"
            "- The edges of the image must remain stable with minimal changes\n"
            "- The video length should be 10 seconds\n"
            "- The main variations should be concentrated on the central object of the image\n"
            "- Preserve the original style, tone, and composition of the image\n"
            "## Technical Parameters\n"
            "- Output format: MP4\n"
            "- Resolution: same as the input image\n"
            "- Frame rate: 24fps\n"
            "- Duration: 10 seconds\n"
            "## Types of Variations\n"
            "- Subtle changes in lighting and shadows\n"
            "- Minor dynamic effects on objects\n"
            "- Slight variations in material appearance\n"
            "- Gentle fluctuations in ambient atmosphere\n"
            "## Prohibited Actions\n"
            "- No large-scale object movements\n"
            "- No camera zoom, pan, or tilt\n"
            "- No alteration of the original composition\n"
            "- No addition of new objects or elements"
        ),
        (
            "human",
            "Please convert this image into a 10-second static-camera video with the following requirements: "
            "1) The camera must remain completely still, as if mounted on a tripod; "
            "2) The animation speed should be smooth and consistent; "
            "3) The frame must stay stable, with changes focused mainly on the central object: {image_description}"
        ),
    ]
)

\end{lstlisting}

\end{document}